\documentclass[pra,reprint,amssymb,amsmath,floatfix,aps,showpacs]{revtex4-1}

\usepackage{graphicx}
\usepackage{braket}
\usepackage{float}
\usepackage{multirow}

\usepackage[usenames,dvipsnames]{color}

\begin{document}

\title{Feshbach resonances, molecular bound states and \\
prospects of ultracold molecule formation in mixtures of ultracold K and Cs}

\author{Hannah J. Patel$^{1}$, Caroline L. Blackley$^{1}$, Simon L. Cornish$^{2}$ and
Jeremy M. Hutson$^{1}$}

\affiliation{$^{1}$Joint Quantum Centre (JQC) Durham/Newcastle, Department of
Chemistry, Durham University, Durham DH1 3LE, United Kingdom}
\affiliation{$^{2}$Joint Quantum Centre (JQC) Durham/Newcastle, Department of
Physics, Durham University, Durham DH1 3LE, United Kingdom}

\date{\today}

\begin{abstract}
We consider the possibilities for producing ultracold mixtures of K and Cs and
forming KCs molecules by magnetoassociation. We carry out coupled-channel
calculations of the interspecies scattering length for $^{39}$KCs, $^{41}$KCs
and $^{40}$KCs and characterize Feshbach resonances due to $s$-wave and
$d$-wave bound states, with widths ranging from below 1~nG to 5~G. We also
calculate the corresponding bound-state energies as a function of magnetic
field. We give a general discussion of the combinations of intraspecies and
interspecies scattering lengths needed to form low-temperature atomic mixtures
and condensates and identify promising strategies for cooling and molecule
formation for all three isotopic combinations of K and Cs.
\end{abstract}


\maketitle

\section{Introduction}
\label{Intro}

There is great current interest in producing ultracold polar molecules
\cite{Carr:NJPintro:2009,Friedrich2009}. Polar molecules can be oriented by
electric fields and then have strongly anisotropic long-range interactions,
which can produce a range of novel quantum phases and opportunities for quantum
simulation and quantum information processing
\cite{Capogrosso-Sansone2010,Micheli:2007,Wall2009}.

One way to produce ultracold molecules is by magnetoassociation of ultracold
atoms followed by laser-induced transfer to a low-lying vibrational level.
Pairs of atoms are first converted into ``Feshbach molecules" in very high
vibrational states by tuning an applied magnetic field across a zero-energy
Feshbach resonance, and the molecules are then transferred into a low-lying
level by stimulated Raman adiabatic passage (STIRAP). Ni {\em et al.}\
\cite{Ni:KRb:2008} have produced $^{40}$K$^{87}$Rb molecules in their absolute
ground state by this route. The ground-state KRb molecules can be transferred
between hyperfine states using microwave radiation
\cite{Ospelkaus:hyperfine-control:2010} and confined in one-dimensional
\cite{deMiranda:2011} and three-dimensional \cite{Chotia:2012} optical
lattices. Non-polar Cs$_2$ \cite{Danzl:ground:2010} and triplet Rb$_2$
\cite{Lang:ground:2008} have been prepared using similar methods.

Many of the alkali-metal dimers, including KRb, can undergo exothermic
bimolecular reactions to form pairs of homonuclear molecules. For fermionic
$^{40}$K$^{87}$Rb, these reactions are suppressed at very low temperatures for
samples of molecules that are all in the same hyperfine level, but proceed very
fast if more than one level is populated \cite{Ospelkaus:react:2010}. Even when
suppressed, the reaction provides a loss mechanism for the ground-state
molecules, and for bosonic molecules no suppression is expected. However,
\.Zuchowski and Hutson \cite{Zuchowski:trimers:2010} have shown that NaK, NaRb,
NaCs, KCs and RbCs are energetically stable to all possible 2-body reactions.
There is therefore particular interest in forming ultracold polar molecules of
these species. Formation of Feshbach molecules has now been achieved for
$^{23}$Na$^{40}$K \cite{Wu:2012} and $^{87}$RbCs \cite{Takekoshi:RbCs:2012,
Koppinger:RbCs:2014}. Takekoshi {\em et al.}\ \cite{Takekoshi:RbCs:2014} have
recently succeeded in producing around 1500 $^{87}$RbCs molecules in their
rovibrational ground state by magnetoassociation followed by STIRAP.

The purpose of the present paper is to investigate the feasibility of producing
KCs molecules by magnetoassociation. KCs is of particular interest because it
is predicted to have a dipole moment of 1.92~D \cite{Aymar:2005}, which is
about 50\% larger than RbCs; this allows dipolar interactions to dominate van
der Waals interactions at lower electric fields, and will make it easier to
suppress collisions that sample short-range effects in quasi-2D geometries
\cite{Julienne:dipoles:2011}. Interaction potentials for the lowest singlet and
triplet states of KCs have been obtained by Ferber {\em et al.}\
\cite{Ferber:2009}, by fitting to extensive electronic spectra in a heat pipe
\cite{Ferber:2008, Ferber:2009}, and have recently been refined to include
coupled levels nearer dissociation \cite{Ferber:2013}. Ferber {\em et al.}\
\cite{Ferber:2009, Ferber:2013} carried out scattering calculations to identify
Feshbach resonances in an $s$-only basis set (limited to functions with $L=0$,
where $L$ is the end-over-end angular momentum of the two atoms about one
another). However, for both RbCs and Cs$_2$, the resonances that have been used
for molecule production arise from bound states with $L=2$ (or higher for
Cs$_2$). In the present work, we therefore carry out scattering and bound-state
calculations including $L=2$ functions, in order to understand the full range
of possibilities. We then give a general discussion of the combinations of
intraspecies and interspecies scattering lengths needed to form low-temperature
atomic mixtures and condensates, and examine the scattering properties to
identify promising strategies for cooling and molecule formation for all three
isotopic combinations of K and Cs.

\section{Theoretical and computational methods}
\label{Theory}

The Hamiltonian for the interaction of two alkali-metal atoms in their ground
$^2S$ states may be written
\begin{equation}
\frac{\hbar^2}{2\mu} \left[-R^{-1} \frac{d^2}{dR^2} R + \frac{\hat
L^2}{R^2} \right] + \hat h_1 + \hat h_2 + \hat V(R),
\label{eq:SE}
\end{equation}
where $\hat L^2$ is the operator for the end-over-end angular momentum of the
two atoms about one another, $\hat h_1$ and $\hat h_2$ are the monomer
Hamiltonians, including hyperfine couplings and Zeeman terms, and $\hat V(R)$
is the interaction operator.

In the present work we solve the scattering and bound-state problems by
coupled-channels calculations using the MOLSCAT \cite{molscat:v14} and BOUND
\cite{Hutson:bound:1993} packages, as modified to handle magnetic fields
\cite{Gonzalez-Martinez:2007}. Both scattering and bound-state calculations use
propagation methods and do not rely on basis sets in the interatomic distance
coordinate $R$. The methodology is exactly the same as described for Cs in
Section IV of Ref.\ \cite{Berninger:Cs2:2013}, so will not be repeated here.
The calculations are performed using a fully uncoupled basis set,
\begin{equation}
|s_1 m_{s1}\rangle|i_1 m_{i1}\rangle
|s_2 m_{s2}\rangle |i_2 m_{i2}\rangle |LM_L\rangle, \label{eqbasdecoup}
\end{equation}
symmetrized for exchange symmetry when the two atoms are identical. $s$ and $i$
are the electron and nuclear spins, respectively. $^{39}$K and $^{41}$K have
$i=3/2$, while $^{40}$K has $i=4$ and an inverted hyperfine structure. The
matrix elements of the different terms in the Hamiltonian in this basis set are
given in the Appendix of Ref.~\cite{Hutson:Cs2-note:2008}. The only rigorously
conserved quantities are the parity, $(-1)^L$, and the projection of the total
angular momentum, $M_{\rm tot}=M_F+M_L$, where
$M_F=m_{s1}+m_{i1}+m_{s2}+m_{i2}$. $M_F$ itself is nearly conserved except near
avoided crossings. The basis sets include all functions for $L=0$ and $L=2$
with the required $M_{\rm tot}$.

The energy-dependent $s$-wave scattering length $a(k)$ is obtained from the
diagonal S-matrix element in the incoming channel,
\begin{equation}
\label{theory:eq5}
a(k) = \frac{1}{ik} \left(\frac{1-S_{00}}{1+S_{00}}\right),
\end{equation}
where $k^2=2\mu E/\hbar^2$ and $E$ is the kinetic energy
\cite{Hutson:res:2007}. Feshbach resonances are initially located using the
FIELD package \cite{Hutson:field:2011}, which provides a complete list of the
magnetic fields at which bound states cross threshold (or cross a specified
energy). The resonances are then characterized by running MOLSCAT at fields
close to resonance, and converging numerically on the pole position using the
formula \cite{Moerdijk:1995}
\begin{equation}
a(B)=a_{\rm bg}\left(1-\frac{\Delta}{B-B_0}\right),
\end{equation}
where $B_0$ is the resonance position and $\Delta$ is its width. This procedure
is able to locate and characterize resonances with widths as small as a few pG.

The interaction operator $\hat V(R)$ may be written
\begin{equation}
{\hat V}(R) = \hat V^{\rm c}(R) + \hat V^{\rm d}(R).
\label{eq:V-hat}
\end{equation}
Here $\hat V^{\rm c}(R)=V_0(R)\hat{\cal{P}}^{(0)} + V_1(R)\hat{\cal{P}}^{(1)}$
is an isotropic potential operator that depends on the electronic potential
energy curves $V_0(R)$ and $V_1(R)$ for the lowest singlet and triplet states
of KCs. The singlet and triplet projectors $\hat{ \cal{P}}^{(0)}$ and $\hat{
\cal{P}}^{(1)}$ project onto subspaces with total electron spin quantum numbers
0 and 1 respectively. The potential curves for the singlet and triplet states
of KCs are taken from Ferber {\em et al.}\ \cite{Ferber:2013}. These curves are
expected to produce resonance positions with an absolute uncertainty of 5 to
10~G, with considerably lower uncertainties in the relative positions. The
potentials for K$_2$ were taken from Falke {\em et al.} \cite{Falke:2008},
while the Cs intraspecies scattering length was taken from the tabulation of
Berninger {\em et al.} \cite{Berninger:Cs2:2013}.

At long range, the coupling $\hat V^{\rm d}(R)$ of Eq.\ \ref{eq:V-hat} has a
simple magnetic dipole-dipole form that varies as $1/R^3$~\cite{Stoof:1988,
Moerdijk:1995}. However, for heavy atoms it is known that second-order
spin-orbit coupling provides an additional contribution that has the same
tensor form as the dipole-dipole term. This contribution dominates at short
range for species containing Cs \cite{Mies:1996, Kotochigova:2001} and has a
large effect on the widths of resonances due to states with $L>0$. In the
present work, $\hat V^{\rm d}(R)$ is represented as
\begin{equation}
\label{eq:Vd} \hat V^{\rm d}(R) = \lambda(R) \left ( \hat s_1\cdot
\hat s_2 -3 (\hat s_1 \cdot \vec e_R)(\hat s_2 \cdot \vec e_R)
\right ) \,,
\end{equation}
where $\vec e_R$ is a unit vector along the internuclear axis and $\lambda$ is
an $R$-dependent coupling constant. For both Cs$_2$ \cite{Kotochigova:2001} and
RbCs \cite{Takekoshi:RbCs:2012}, electronic structure calculations showed that
the second-order spin-orbit splitting can be represented by a biexponential
form, so that the overall form of $\lambda(R)$ is
\begin{eqnarray}
\label{eq:lambda}
\lambda(R) &=& E_{\rm h} \alpha^2 \bigg[
A_{\rm 2SO}^{\rm short} \exp\left(-\beta_{\rm 2SO}^{\rm short}(R/a_0)\right)
\nonumber\\
&+& A_{\rm 2SO}^{\rm long}
\exp\left(-\beta_{\rm 2SO}^{\rm long}(R/a_0)\right)
+  \frac{1}{(R/a_0)^3}\bigg],
\end{eqnarray}
where $\alpha\approx 1/137$ is the atomic fine-structure constant and $a_0$ is
the Bohr radius. The second-order spin-orbit coupling has not been calculated
for KCs, but an estimate may be made from the values for Cs$_2$ and RbCs. It is
physically reasonable to suppose that the coupling comes principally from the
Cs atom(s) and (for chemically similar species) does not depend strongly on the
identity of the other atom. Evaluating the second-order spin-orbit contribution
to $\lambda(R)$ \cite{Kotochigova:2001, Takekoshi:RbCs:2012} at the inner
turning point of the triplet curve at zero energy gives values per Cs atom
within about 40\% of one another for Cs$_2$ and RbCs. In the present work, we
simply shifted the RbCs function inwards 0.125 $a_0$, to give the same value at
the inner turning point for KCs as for RbCs. We thus use $\beta_{\rm 2SO}^{\rm
short} = 0.80$ and $\beta_{\rm 2SO}^{\rm long} = 0.28$, as for RbCs
\cite{Takekoshi:RbCs:2012}, with $A_{\rm 2SO}^{\rm short} = -45.5$ and $A_{\rm
2SO}^{\rm long} = -0.032$.

\section{Results}
\subsection{$^{39}$KCs}

\begin{figure*}[t]
\centering
\includegraphics[width=0.95\textwidth]{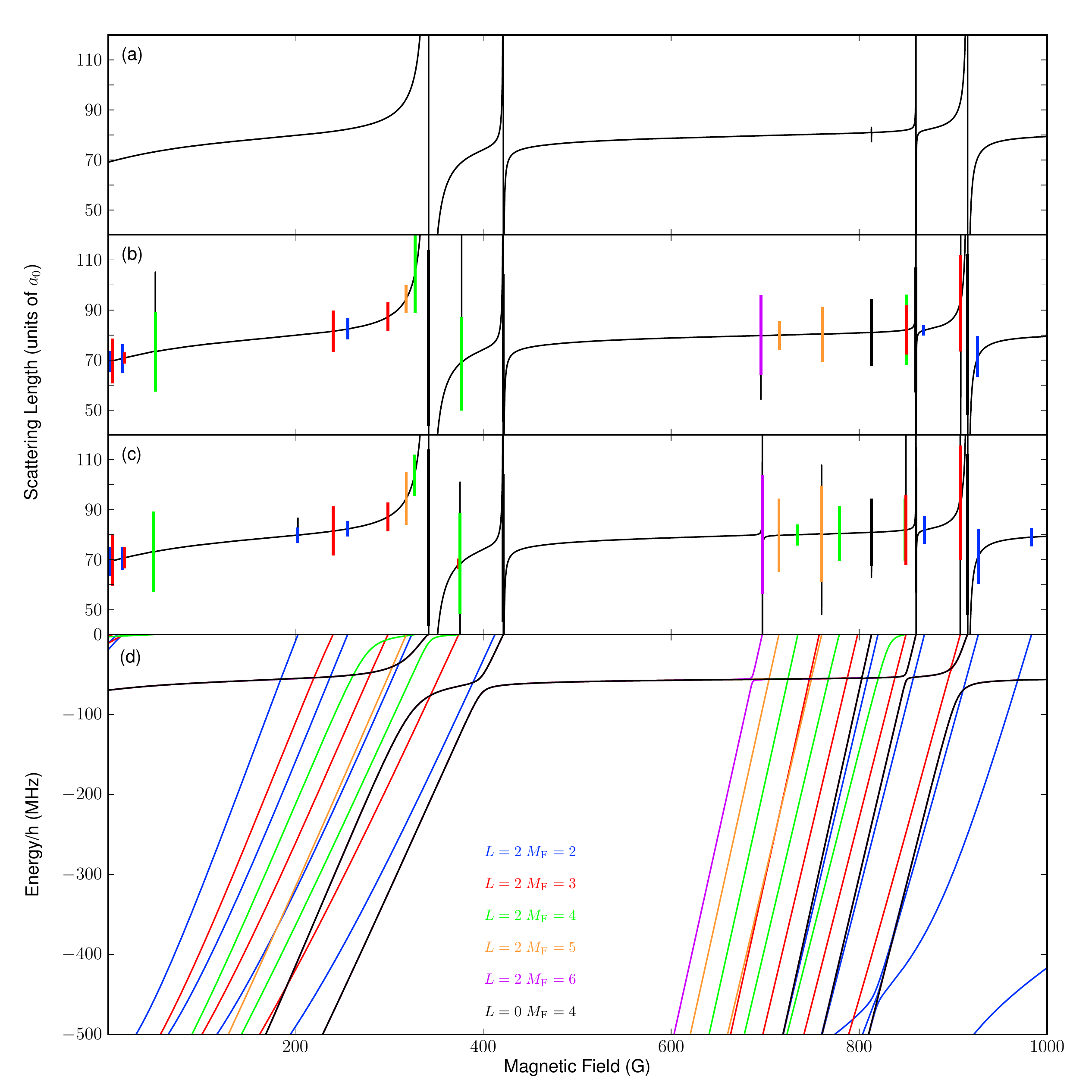}
\caption{(Color online) $^{39}$K$^{133}$Cs: (a) $L=0$ functions only; (b) $L=0$
and 2 functions, but 2nd-order spin-orbit coupling not included; (c) scattering
length with $L=0$ and 2 functions, and approximate model of 2nd-order
spin-orbit coupling included; (d) bound states. Resonance widths greater than
1~$\mu$G are shown as vertical bars with lengths proportional to $\log
\Delta/\mu$G.} \label{fig:K39}
\end{figure*}

Figure \ref{fig:K39} shows the scattering length and the corresponding
near-threshold bound states for $^{39}$KCs. When only $s$-wave basis functions
are included (Fig.\ \ref{fig:K39}(a)), the scattering length shows 5 resonances
below 1000~G, where $s$-wave bound states (black in Fig.\ \ref{fig:K39}(d))
cross threshold. These agree within 1~G with those calculated by Ferber {\em et
al.}\ \cite{Ferber:2013}. When $d$-wave ($L=2$) basis functions are included,
an additional 30 bound states cross threshold below 1000~G. These are
color-coded according to $M_F$ in Fig.\ \ref{fig:K39}(d). If only the
long-range spin-spin coupling is included (the $R^{-3}$ term in Eq.\
\ref{eq:lambda}), the resonances due to $d$-wave states are quite narrow (Fig.\
\ref{fig:K39}(b)). However, if second-order spin-orbit coupling is included,
most of them become significantly broader, as shown in Fig.\ \ref{fig:K39}(c)
(note the logarithmic scale of the vertical bars used to indicate the resonance
widths). Some of the resonances have widths suitable for use in molecule
formation. The positions and widths of some of the broader resonances are given
in Table \ref{table:AllKCsFeshbachResonances}, and a complete tabulation (all
resonances, with and without second-order spin-orbit coupling for $^{39}$KCs)
is included in Supplemental Material \footnote{See Supplemental Material at [to
be inserted by publisher] for a full listing of the resonance positions and
widths.}.

It is clearly important to include second-order spin-orbit coupling in
calculations involving heavy atoms such as Cs. In the following sections, only
calculations including the second-order spin-orbit coupling will be presented.

\begin{table}[!t]
\begin{center}
\begin{tabular}{ccccc}
\hline
\hline
 $B_0$ (G)&  $\Delta$ (G)& $a_{\rm{bg}}$~($a_0$) & $L$ & $M_{F}$  \\
\hline
\multicolumn{5}{c}{$^{39}$KCs}		\\
\hline	
49.57	&$	0.001	$&$	73.2	$&$	2	$&$	4	$ \\
341.90	&$	4.8	$&$	79.0	$&$	0	$&$	4	$ \\
375.35	&$	0.006	$&$	68.5	$&$	2	$&$	4	$ \\
421.36	&$	0.4	$&$	74.7	$&$	0	$&$	4	$ \\
697.02	&$	0.03	$&$	80.0	$&$	2	$&$	6	$ \\
760.13	&$	0.004	$&$	80.3	$&$	2	$&$	5	$ \\
813.14	&$	3 \times 10^{-4}	$&$	81.0	$&$	0	$&$	4	$ \\
860.52	&$	0.05	$&$	82.0	$&$	0	$&$	4	$ \\
907.54	&$	0.02	$&$	92.7	$&$	2	$&$	3	$ \\
915.56	&$	1.2	$&$	80.1	$&$	0	$&$	4	$ \\
\hline
\multicolumn{5}{c}{$^{40}$KCs}		\\
\hline						
57.59	&$	< 10^{-9}$&$	-40.3	$&$	0	$&$	- 3/2	$ \\
69.85	&$	< 10^{-9}$&$	-40.3	$&$	0	$&$	- 3/2	$ \\
89.01	&$	< 10^{-9}$&$	-40.3	$&$	0	$&$	- 3/2	$ \\
122.77	&$	< 10^{-9}$&$	-40.3	$&$	0	$&$	- 3/2	$ \\
192.18	&$	-0.001	$&$	-40.2	$&$	2	$&$	- 3/2	$ \\
196.71	&$	-3 \times 10^{-7}	$&$	-40.2	$&$	0	$&$	- 3/2	$ \\
215.96	&$	-0.01	$&$	-40.2	$&$	2	$&$	- 1/2	$ \\
230.24	&$	< 10^{-9}$&$-40.2$&$	0	$&$	- 3/2	$ \\
234.15	&$	< 10^{-9}$&$	-40.1	$&$	0	$&$	- 3/2	$ \\
239.55	&$	< 10^{-9}$&$	-40.1	$&$	0	$&$	- 3/2	$ \\
246.44	&$	-4 \times 10^{-7}	$&$	-40.0	$&$	0	$&$	- 3/2	$ \\
254.52	&$	-1 \times 10^{-4}	$&$	-39.8	$&$	0	$&$	- 3/2	$ \\
264.34	&$	-0.1 	$&$	-40.3	$&$	0	$&$	- 3/2	$ \\
379.60	&$	-0.002 	$&$	-40.3	$&$	2	$&$	- 5/2	$ \\
470.25	&$	-0.01	$&$	-40.2	$&$	0	$&$	- 3/2	$ \\
677.44	&$	< 10^{-9}$&$	-40.2	$&$	0	$&$	- 3/2	$ \\
902.84	&$	< 10^{-9}$&$	-40.2	$&$	0	$&$	- 3/2	$ \\
\hline
\multicolumn{5}{c}{$^{41}$KCs}		\\
\hline
23.89	&$	0.02	$&$	193.0	$&$	2	$&$	6	$ \\
25.68	&$	0.03	$&$	189.3	$&$	2	$&$	5	$ \\
28.41	&$	0.007	$&$	188.7	$&$	2	$&$	4	$ \\
87.38	&$	0.003	$&$	201.6	$&$	2	$&$	4	$ \\
90.10	&$	0.008	$&$	201.1	$&$	2	$&$	5	$ \\
94.28	&$	0.001	$&$	201.5	$&$	2	$&$	3	$ \\
109.86	&$	0.002	$&$	204.5	$&$	2	$&$	2	$ \\
120.89	&$	0.02	$&$	206.0	$&$	0	$&$	4	$ \\
168.19	&$	0.6	$&$	262.6	$&$	0	$&$	4	$ \\
171.20	&$	1.2	$&$	151.3	$&$	0	$&$	4	$ \\
861.03	&$	0.03	$&$	247.7	$&$	2	$&$	5	$ \\
884.92	&$	4.1	$&$	211.4	$&$	0	$&$	4	$ \\
966.89	&$	0.1 $&$	201.5	$&$	0	$&$	4	$ \\
\hline
\hline
\end{tabular}
\end{center}
\caption{Listing of all $s$-wave Feshbach resonances and $d$-wave Feshbach
resonances with widths over 1~mG for all isotopes of KCs in the field range 0
to 1000~G. A complete tabulation of all resonances is included in Supplemental
Material.} \label{table:AllKCsFeshbachResonances}
\end{table}

\subsection{$^{41}$KCs}

\begin{figure*}[t]
\centering
\includegraphics[width=0.95\textwidth]{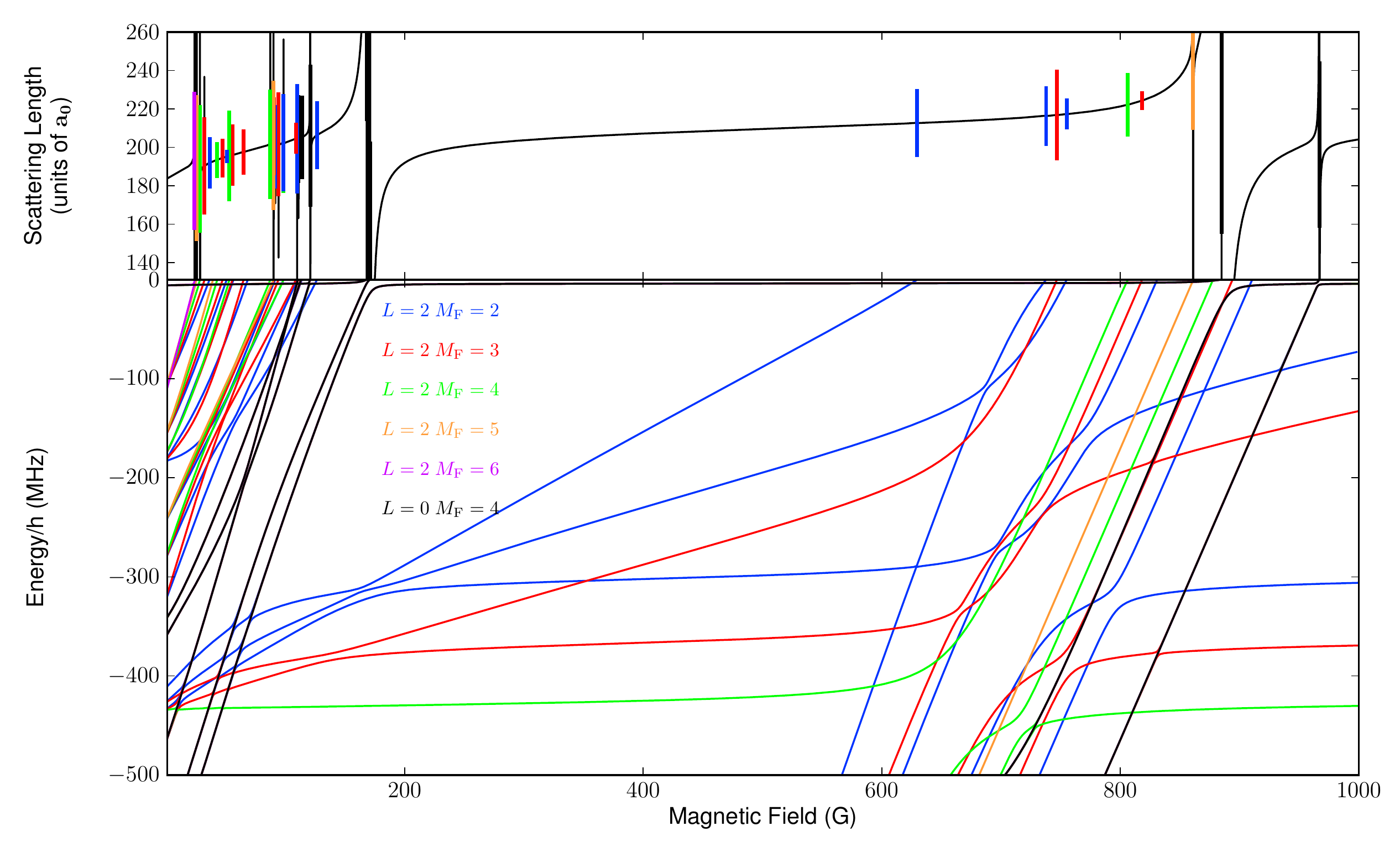}
\caption{(Color online) $^{41}$K$^{133}$Cs: (a) scattering length with $L=0$
and 2 functions, and approximate model of 2nd-order spin-orbit coupling; (b)
bound states. Resonance widths greater than 1~$\mu$G are shown as vertical bars
with lengths proportional to $\log_{10} \Delta/\mu$G, on the same scale as in
Fig.\ \ref{fig:K39}.} \label{fig:K41}
\end{figure*}

Figure \ref{fig:K41} shows the scattering length and the corresponding
near-threshold bound states for $^{41}$KCs. In this case there is an $s$-wave
bound state only about 3~MHz $\times h$ below threshold, which produces a
background scattering length that is large and positive (though not so large as
for $^{87}$RbCs and Cs$_2$). There are 7 resonances below 1000~G due to
$s$-wave states; it may be noted that Ferber {\em et al.}\ \cite{Ferber:2009}
were able to locate only 6 of these. In addition, the rich set of $d$-wave
levels produces 24 resonances below about 130~G, some of which offer good
prospects for molecule formation, as discussed below.

\subsection{$^{40}$KCs}

\begin{figure*}[t]
\centering
\includegraphics[width=0.95\textwidth]{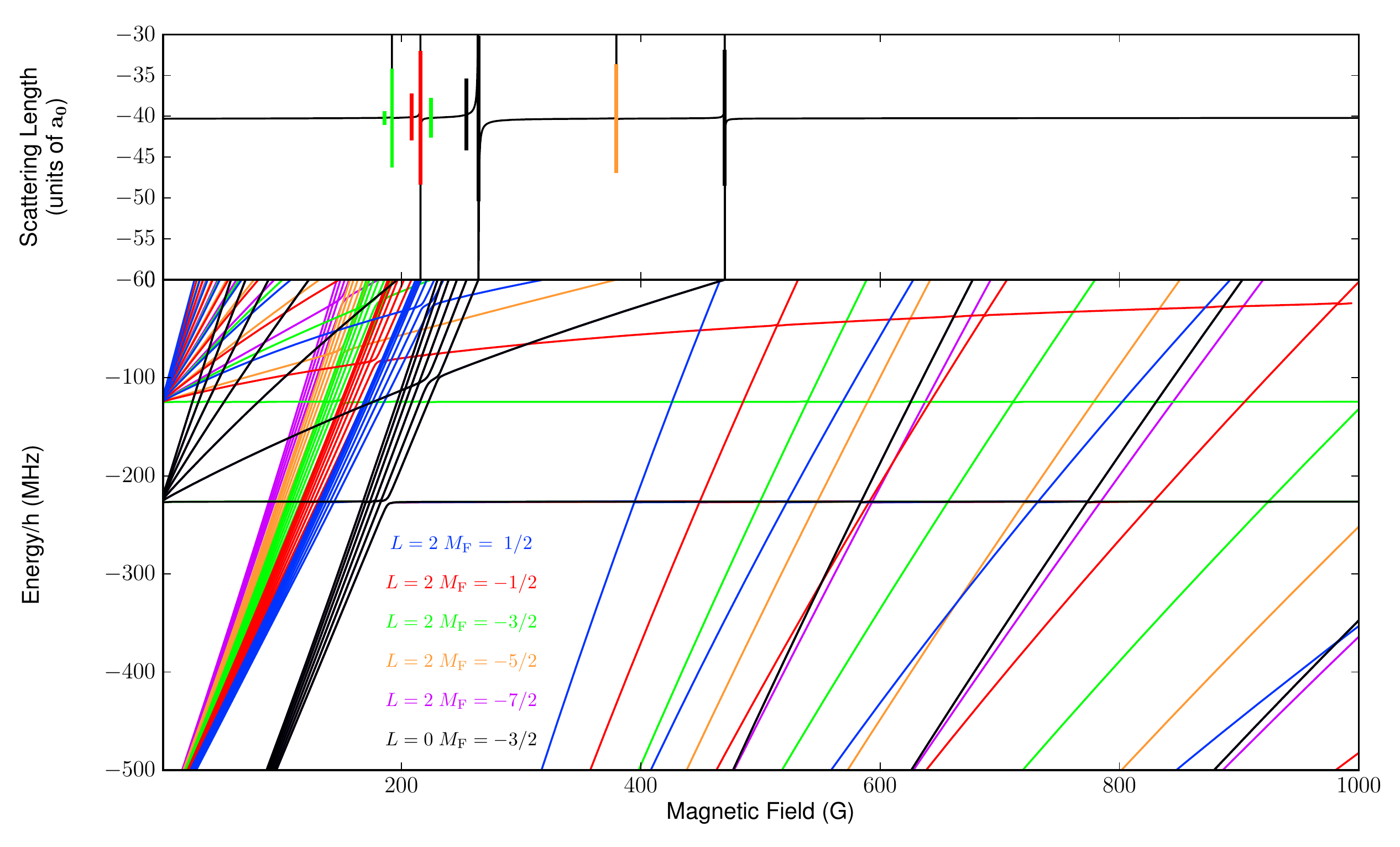}
\caption{(Color online) $^{40}$K$^{133}$Cs: (a) scattering length with $L=0$
and 2 functions, and approximate model of 2nd-order spin-orbit coupling; (b)
bound states. Resonance widths greater than 1~$\mu$G are shown as vertical bars
with lengths proportional to $\log_{10} \Delta/\mu$G, on the same scale as in
Fig.\ \ref{fig:K39}.} \label{fig:K40}
\end{figure*}

Figure \ref{fig:K40} shows the scattering length and the corresponding
near-threshold bound states for $^{40}$KCs. In this case there is a very dense
bound-state spectrum, with 14 $s$-wave levels and a further 70 $d$-wave levels
crossing threshold below 1000~G. However, the resulting resonances are all
quite weak (note the contracted vertical scale of Fig.\ \ref{fig:K40}). Indeed,
Ferber {\em et al.}\ \cite{Ferber:2009} located only the 2 broadest resonances,
although they commented that a further 13 $s$-wave resonances should exist. The
widest $d$-wave resonance has $\Delta\approx 10$~mG, and many have $\Delta <
1$~nG. The narrowness arises because the singlet and triplet scattering lengths
are quite similar, which directly reduces the strength of the resonances due to
$s$-wave states \cite{Julienne:1997} and indirectly reduces the strength of
those due to $d$-wave states as well.

\section{Prospects for producing KC\lowercase{s} molecules}

We now discuss the implications of the predicted scattering lengths and
interspecies Feshbach resonances for the formation of mixtures of ultracold K
and Cs and the production of ultracold KCs molecules. We begin with a general
discussion of the scattering properties required, and then proceed to consider
their implications for the isotopologs of KCs.

It should be noted that the remaining uncertainties in the KCs potential curves
of ref.\ \cite{Ferber:2013} may shift the interspecies resonances by a few
Gauss, and may thus change the quantitative predictions in some cases, but the
overall picture is robust. Once resonance positions are observed for one
isotopic combination, it will be possible to refine the potentials and improve
the detailed predictions.

\subsection{General considerations}

Magnetoassociation has been demonstrated in a wide range of atomic systems,
using resonances with widths varying from a few mG to over 10~G
\cite{ChinRMP2010}. The wider resonances allow slower magnetic field ramp
speeds whilst maintaining adiabaticity through the avoided crossing between
atomic and molecular states. However, dwelling too long near the pole of the
resonance can lead to enhanced 3-body losses and in practice the speed of the
magnetic field ramp has to be carefully optimized. The overall conversion
efficiency is dictated by the phase-space density of the atomic gas.
Magnetoassociation can be carried out in a thermal gas close to degeneracy
\cite{HodbyPRL2005}, but it is best performed in the quantum-degenerate regime;
this produces a molecular gas that is colder and has higher phase-space
density, so is more suitable for experiments that aim for quantum coherence.

The need to produce an atomic gas close to quantum degeneracy for
magnetoassociation requires the ratio of ``good'' (elastic) to ``bad''
(inelastic and 3-body loss) collisions to be suitable for efficient evaporative
cooling \cite{KetterleAdv1996,WeinerRMP1999}. Whilst loss rates vary
considerably from one species to another, it is generally observed in
single-species experiments that the evaporation is most efficient if the
$s$-wave scattering length $a$ satisfies $40\ a_0 \lesssim |a| \lesssim 250\
a_0$ \cite{AndersonScience1995,DavisPRL1995,TakusuPRL2003,KishimotoPRA2009}.
Below this range the elastic collision rate (which scales as $a^2$ in the
ultracold limit) becomes too low  for cooling to proceed efficiently
\footnote{For lighter species such as $^7$Li \cite{BradleyPRL1997} and
metastable He \cite{RobertScience2001,SantosPRL2001}, slightly smaller
scattering lengths can be used because of the inverse scaling of the collision
rate with mass.}, and above it the 3-body loss rate (which scales as $a^4$
\cite{BedaquePRL2000}) becomes too high. For species where the background
scattering length falls outside this range, magnetic tuning of the scattering
length near Feshbach resonances has proved an essential tool to reach quantum
degeneracy
\cite{CornishPRL2000,StreckerNature2002,WeberScience2003,LandiniPRA2012}. In
such cases, the inelastic and 3-body loss rates also tune with magnetic field
\cite{RobertsPRL2000} allowing the ratio of elastic collisions to those causing
loss to be optimized \cite{MarchantPRA2012}. In some cases, most notably Cs,
the $a^4$ scaling of the 3-body loss rate is significantly modified by the
Efimov effect \cite{EfimovPLB1970,BraatenPhysRep2006}, allowing particularly
efficient evaporative cooling around the resulting 3-body recombination minimum
\cite{Kraemer:2006,Berninger:Efimov:2011,Berninger:Cs2:2013}.

In the case of an atomic mixture, the need to cool both species to high
phase-space densities puts conditions on both intraspecies scattering lengths
(hereafter $a_{11}$ and $a_{22}$) and the interspecies scattering length
($a_{12}$). There are several different scenarios that can produce efficient
cooling. If $a_{11}$ and $a_{12}$ are in the desired range, species 1 can be
cooled directly and species 2 will be cooled by interspecies collisions
\cite{MyattPRL1997,ModugnoPRL2002,HadzibabicPRL2002,RoatiPRL2002,SilberPRL2005,PappPRL2008}.
In this case the only restriction on $a_{22}$ is that its magnitude should not
be so large as to produce unacceptable 3-body loss. This typically restricts
$|a_{22}|$ to values below about 600~$a_0$. Alternatively, if $a_{11}$ and
$a_{22}$ are in the desired range but $a_{12}$ is not, the two species can be
cooled independently, either in the same trap (if $|a_{12}|$ is small
\cite{ChoPRA2013}) or separately (if $|a_{12}|$ is large
\cite{Takekoshi:RbCs:2012}). In the latter case it is necessary to overlap the
clouds in a subsequent step. If none of the above scenarios can be realized for
atoms in their absolute ground states, it may be possible to cool one or both
species in an excited Zeeman or hyperfine state, although this introduces the
possibility of further loss via 2-body inelastic collisions. Finally, if the
desired conditions cannot be fulfilled at a single magnetic field, it might be
feasible to cool the two species in separate traps at different fields, though
this would introduce substantial extra complexity.

If cooling proceeds all the way to Bose-Einstein condensation, there are
further restrictions. Large individual condensates are stable with respect to
collapse only if $a_{11}>0$ and $a_{22}>0$, and the mixed condensate requires
in addition that $g_{12}^2<g_{11}g_{22}$, where the interaction coupling
constants are \cite{RiboliPRA2002}
\begin{equation}
g_{ij} = 2 \pi \hbar^{2} a_{ij}\left(\frac{m_{i}+m_{j}}{m_{i} m_{j}}\right).
\end{equation}
If $g_{12}$ is too positive, the mean-field repulsion leads to phase
separation, whereas if it is too negative the mixed condensate collapses.
However, the magnetic field at which the condensates are mixed need not be the
same as the one used for the early stages of evaporative or sympathetic cooling
\cite{Koppinger:RbCs:2014}. The instabilities can also, in principle, be
avoided by loading the two species into a 3-dimensional optical lattice prior
to molecule formation to realize a Mott-insulator phase with exactly one atom
of each species per lattice site \cite{DamskiPRL2003}.

Magnetoassociation can in principle be carried out at a field different from
that used to form the atomic mixture. However, the mixture must be stable for
the duration of the magnetoassociation sequence, which is typically a few
milliseconds. For a mixed condensate, the stability conditions are those given
above. For a thermal gas close to degeneracy, the conditions are less
restrictive but magnetoassociation is less efficient.

\begin{figure*}[t]
\includegraphics[width=0.95\textwidth]{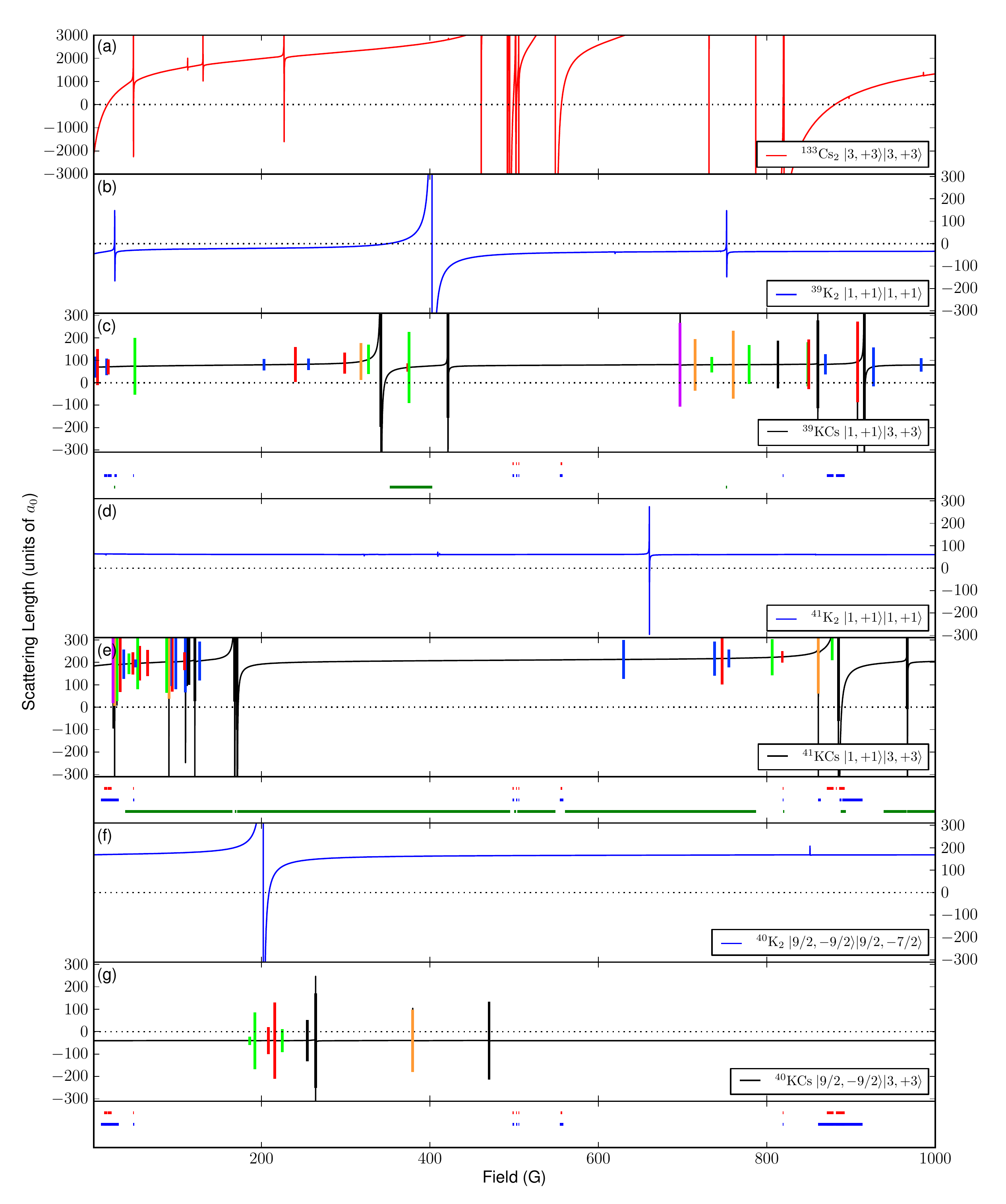}
\caption{(Color online) Scattering lengths for isotopologs of KCs, together
with those of the corresponding isotopes of K and Cs. The scattering lengths
are shown with the same magnetic field axis, to facilitate the identification
of regions where the combination is conducive to molecule formation. Resonance
widths greater than 1~$\mu$G are shown as vertical bars with lengths
proportional to $\log_{10} \Delta/\mu$G. The colored bars beneath each
interspecies scattering length indicate the fields at which both species can be
cooled evaporatively (red, top), the fields at which one species can be cooled
evaporatively and the other sympathetically (blue, center), and the fields at
which the condensates are miscible (green, bottom, not shown for $^{40}$K).}
\label{a-allspecies}
\end{figure*}

\subsection{Implications for KCs}

Figure \ref{a-allspecies} shows the intraspecies and interspecies scattering
lengths for all the atoms important for KCs molecule formation, together with
the positions of interspecies resonances. The colored bars below each
interspecies scattering length show the fields at which both species can be
cooled evaporatively (red, top), the fields at which one species can be cooled
evaporatively and the other sympathetically (blue, center), and the fields at
which the condensates are stable and miscible (green, bottom).

The intraspecies scattering length for Cs is shown in Fig.\
\ref{a-allspecies}(a). The large background value places severe limitations on
the magnetic fields where efficient evaporative cooling is possible.
Bose-Einstein condensates of Cs have been produced only in the $f=3, m_f=3$
ground state, where 2-body inelastic collisions cannot occur. Even in this
state, the magnetic field used for cooling must be chosen to tune the
scattering length to a moderate positive value to prevent excessive 3-body
losses. Condensates of Cs are usually produced at fields around 21~G
\cite{WeberScience2003,HungPRA2008,McCarronPRA2011}, just above the zero
crossing in the scattering length at 17~G. Similar windows of moderate positive
scattering lengths exist above zero crossings at 556~G and 881~G, associated
with broad Feshbach resonances at 549~G and 787~G; Berninger {\em et al.}\
\cite{Berninger:Cs2:2013} carried out cooling at 558.7~G and 894~G, where
3-body losses are low \cite{Berninger:Efimov:2011}.
Table~\ref{table:ScatLength_compare} summarizes the intraspecies and
interspecies scattering lengths at the boundaries of the regions of moderate
positive Cs scattering length.

\begin{table}[!t]
\begin{center}
\begin{tabular}{cccccccc}
\hline
\hline
& \multicolumn{7}{c}{Scattering length ($a_0$)}\\
Field (G)  	& $^{133}$Cs 		& 	$^{39}$K 	& 	$^{39}$KCs 	& 	$^{41}$K	 & 	$^{41}$KCs 	& $^{40}$K 		& $^{40}$KCs	 \\
\hline
17.7	&$	35.2	$&$	-34.0	$&$	70.8	$&$	63.1	$&$	189.6	$&$	169.6	$&$	-40.3	$\\
21.7	&$	252.5	$&$	-30.7	$&$	71.1	$&$	62.9	$&$	192.2	$&$	169.9	$&$	-40.3	$\\
556.2	&$	28.2	$&$	-40.0	$&$	78.2	$&$	60.5	$&$	210.9	$&$	165.7	$&$	-40.3	$\\
556.9	&$	253.6	$&$	-40.0	$&$	78.2	$&$	60.5	$&$	210.9	$&$	165.8	$&$	-40.3	$\\
882.3	&$	39.3	$&$	-34.5	$&$	83.1	$&$	60.3	$&$	
539.9$\footnote{$^{41}$KCs has a resonance at 884.9~G that substantially affects its scattering length in this region.}	&$	167.8	$&$	-40.2	$\\
892.2	&$	251.3	$&$	-34.5	$&$	84.5	$&$	60.3	$&$	93.2^{\rm a}	$&$	167.8	$&$	-40.2	$\\
\hline
\hline
\end{tabular}
\end{center}
\caption{Intraspecies and interspecies scattering lengths at fields that bound
the regions where $40\ a_0 \lesssim a_{\rm Cs} \lesssim 250\ a_0$.}
\label{table:ScatLength_compare}
\end{table}

The intraspecies scattering length for $^{39}$K is shown in Fig.\
\ref{a-allspecies}(b). It varies weakly with magnetic field, and is generally
small and negative except near the Feshbach resonance at 402~G. The small value
limits the rate of rethermalization and the negative sign makes a
single-species condensate of $^{39}$K unstable at most fields. These hurdles
have been overcome to produce condensates in the $f=1, m_f=1$ ground state by
using a combination of sympathetic cooling with $^{87}$Rb and Feshbach tuning
of the $^{39}$K scattering length \cite{RoatiPRL2007}. More recently, direct
evaporation of $^{39}$K to condensation has been achieved using several
different Feshbach resonances, including the one at 402~G in the ground state
\cite{LandiniPRA2012}. The interspecies scattering length for $^{39}$KCs shown
is shown in Fig.\ \ref{a-allspecies}(c); at 21~G it is approximately +70~$a_0$,
which may allow sympathetic cooling of $^{39}$K by Cs. The situation is similar
at 558.7~G and 894~G. For $^{39}$KCs, most of the resonances that are broad
enough for magnetoassociation lie above 300~G, at fields where the $^{39}$K
intraspecies scattering length is negative. The Cs intraspecies scattering
length is very large at most $^{39}$KCs resonances that lie below the Cs pole
at 787~G, but crosses zero near 881~G \cite{Berninger:Cs2:2013} and has more
moderate values at fields around the zero crossing. The most promising
$^{39}$KCs resonances for magnetoassociation in a thermal gas are therefore
those at 861~G, 908~G and 916~G, where the Cs scattering length is calculated
to be $-630$, 510 and 630 bohr, respectively. There is also a region from 353
to 402~G, where the condensates are predicted to be stable and miscible. The
resonance at 375~G ($\Delta=6$ mG) is very promising for magnetoassociation to
form $^{39}$KCs in a mixed condensate. Just below 353 G, the $^{39}$K
intraspecies scattering length is small and negative, but it has reached only
$-4\ a_0$ at the $^{39}$KCs resonance at 342~G ($\Delta=4.8$~G). It may
therefore be possible to sweep the magnetic field across this resonance to
perform magnetoassociation before the condensate has time to collapse. The Cs
intraspecies scattering length is very large at these resonances (2540~$a_0$ at
375~G and 2420~$a_0$ at 342~G \cite{Berninger:Cs2:2013}), but
magnetoassociation to form $^{87}$RbCs has been achieved in a thermal gas
\cite{Takekoshi:RbCs:2012, Takekoshi:RbCs:2014, Koppinger:RbCs:2014} at a
resonance at 197.1~G, where the Cs scattering length is 1970~$a_0$
\cite{Berninger:Cs2:2013}, and 3-body collisions are suppressed by a factor of
6 in a condensate \cite{Burt:1997} compared to a thermal gas.

The intraspecies scattering length for $^{41}$K is shown in Fig.\
\ref{a-allspecies}(d). By contrast to $^{39}$K, it varies only slightly from a
background value of approximately $+63~a_0$. Direct evaporation to
Bose-Einstein condensation has been achieved in the magnetically trappable
$f=2, m_f=2$ excited state \cite{KishimotoPRA2009}, where the scattering length
is $+61~a_0$ \cite{Falke:2008}. Efficient evaporation is also expected for
atoms in the $f=1, m_f=1$ ground state confined in an optical dipole trap. The
interspecies scattering length for $^{41}$KCs is shown in Fig.\
\ref{a-allspecies}(e); at 21~G it is approximately 200~$a_0$, which will lead
to rapid interspecies thermalization and may allow $^{41}$K to be used as a
sympathetic coolant for Cs following the approach used with $^{87}$Rb
\cite{McCarronPRA2011}. The situation is similar around 558.7~G and 894~G. At
all these fields, the combination of a moderately large interspecies scattering
length and a relatively low Cs scattering length will produce phase separation
if the mixture is cooled to degeneracy. However, in all cases there are nearby
fields where a larger Cs scattering length makes the condensates miscible.
$^{41}$KCs has a rich spectrum of usable resonances, including 10 resonances
below 200~G. In particular, there are three resonances below 30~G which lie
close to a region where efficient evaporative cooling of both species is
possible. These are comparable in width to the Cs resonance at 19.8~G that has
been used extensively in the creation and study of ultracold Cs$_2$ molecules
\cite{ChinPRL2005,HerbigScience2003,MarkEPL2005}. Since the combination of
scattering lengths at these fields will lead to phase separation, these
resonances would need to be accessed in thermal gases. However, the condensates
are expected to be miscible at most fields between 38 and 786~G and above
939~G, and there are several resonances in these regions that are promising for
magnetoassociation in a mixed condensate. There is also a small region of
miscibility (5~G wide) immediately above the broad $^{41}$KCs resonance at
885~G ($\Delta=4.1$~G), where $a_{\rm K}$ is 63~$a_0$ and $a_{\rm Cs}$ is
around 200~$a_0$. This combination of properties is very promising for cooling
a mixed gas directly to degeneracy, followed by magnetoassociation.

The $s$-wave scattering length is undefined for two $^{40}$K atoms in identical
states because of their fermionic character. Because of this, $^{40}$K is
usually cooled by interspecies collisions between atoms in different Zeeman
states \cite{DeMarcoScience1999}. The scattering length for collisions between
atoms in the $f=9/2, m_f=-9/2$ and $f=9/2, m_f=-7/2$ states is shown in Fig.\
\ref{a-allspecies}(f). It varies weakly with magnetic field, and is generally
around +170~$a_0$ except near the Feshbach resonance at 202~G. The interspecies
scattering length for $^{40}$KCs is approximately $-40\ a_0$ at nearly all
fields. It may therefore be possible to use sympathetic cooling of $^{40}$K
with Cs, in one of the regions where Cs itself can be cooled, to avoid the need
for two different spin states of $^{40}$K. As discussed above, most of the
resonances for $^{40}$KCs are exceedingly narrow. However, resonances that may
be suitable for the production of fermionic molecules are predicted to exist at
192, 216, 264 and 470~G.

\section{Conclusions}

We have explored the possibilities for magnetoassociation to form ultracold KCs
molecules, using calculations of bound states and scattering properties. We
have calculated the interspecies scattering length for $^{39}$KCs, $^{41}$KCs
and $^{40}$KCs. We have characterized Feshbach resonances in $s$-wave
scattering due to $s$-wave and $d$-wave bound states, with widths ranging from
below 1~nG to 5~G, and we have carried out bound-state calculations to identify
the quantum states responsible.

We have considered the combinations of intraspecies and interspecies scattering
lengths to identify possibilities for producing both mixed condensates and
low-temperature thermal mixtures of K and Cs. All the isotopic combinations
offer promising possibilities for producing mixtures with high phase-space
densities and for magnetoassociation to form Feshbach molecules.

Our calculations of interspecies scattering lengths are based on the singlet
and triplet potential curves of ref.\ \cite{Ferber:2013}. Remaining
uncertainties in the potential curves result in uncertainties of at least a few
G in the predicted resonance positions. Once the actual positions of
interspecies Feshbach resonances have been measured and assigned for at least
one isotopolog of KCs, it will be possible to refine the potentials to make
more accurate predictions of the scattering properties.

\medskip\section{Acknowledgments}

The authors acknowledge the support of Engineering and Physical Sciences
Research Council Grant no.\ EP/I012044/1 and EOARD Grant FA8655-10-1-3033. CLB
is supported by a Doctoral Fellowship from Durham University.

\bibliography{../all,../85RbCs/Rb85Cs,KCs_expt}

\begin{thebibliography}{76}%
\makeatletter
\providecommand \@ifxundefined [1]{%
 \@ifx{#1\undefined}
}%
\providecommand \@ifnum [1]{%
 \ifnum #1\expandafter \@firstoftwo
 \else \expandafter \@secondoftwo
 \fi
}%
\providecommand \@ifx [1]{%
 \ifx #1\expandafter \@firstoftwo
 \else \expandafter \@secondoftwo
 \fi
}%
\providecommand \natexlab [1]{#1}%
\providecommand \enquote  [1]{``#1''}%
\providecommand \bibnamefont  [1]{#1}%
\providecommand \bibfnamefont [1]{#1}%
\providecommand \citenamefont [1]{#1}%
\providecommand \href@noop [0]{\@secondoftwo}%
\providecommand \href [0]{\begingroup \@sanitize@url \@href}%
\providecommand \@href[1]{\@@startlink{#1}\@@href}%
\providecommand \@@href[1]{\endgroup#1\@@endlink}%
\providecommand \@sanitize@url [0]{\catcode `\\12\catcode `\$12\catcode
  `\&12\catcode `\#12\catcode `\^12\catcode `\_12\catcode `\%12\relax}%
\providecommand \@@startlink[1]{}%
\providecommand \@@endlink[0]{}%
\providecommand \url  [0]{\begingroup\@sanitize@url \@url }%
\providecommand \@url [1]{\endgroup\@href {#1}{\urlprefix }}%
\providecommand \urlprefix  [0]{URL }%
\providecommand \Eprint [0]{\href }%
\providecommand \doibase [0]{http://dx.doi.org/}%
\providecommand \selectlanguage [0]{\@gobble}%
\providecommand \bibinfo  [0]{\@secondoftwo}%
\providecommand \bibfield  [0]{\@secondoftwo}%
\providecommand \translation [1]{[#1]}%
\providecommand \BibitemOpen [0]{}%
\providecommand \bibitemStop [0]{}%
\providecommand \bibitemNoStop [0]{.\EOS\space}%
\providecommand \EOS [0]{\spacefactor3000\relax}%
\providecommand \BibitemShut  [1]{\csname bibitem#1\endcsname}%
\let\auto@bib@innerbib\@empty
\bibitem [{\citenamefont {Carr}\ \emph {et~al.}(2009)\citenamefont {Carr},
  \citenamefont {{DeMille}}, \citenamefont {Krems},\ and\ \citenamefont
  {Ye}}]{Carr:NJPintro:2009}%
  \BibitemOpen
  \bibfield  {author} {\bibinfo {author} {\bibfnamefont {L.~D.}\ \bibnamefont
  {Carr}}, \bibinfo {author} {\bibfnamefont {D.}~\bibnamefont {{DeMille}}},
  \bibinfo {author} {\bibfnamefont {R.~V.}\ \bibnamefont {Krems}}, \ and\
  \bibinfo {author} {\bibfnamefont {J.}~\bibnamefont {Ye}},\ }\href@noop {}
  {\bibfield  {journal} {\bibinfo  {journal} {New J. Phys.}\ }\textbf {\bibinfo
  {volume} {11}},\ \bibinfo {pages} {055049} (\bibinfo {year}
  {2009})}\BibitemShut {NoStop}%
\bibitem [{\citenamefont {Friedrich}\ and\ \citenamefont
  {Doyle}(2009)}]{Friedrich2009}%
  \BibitemOpen
  \bibfield  {author} {\bibinfo {author} {\bibfnamefont {B.}~\bibnamefont
  {Friedrich}}\ and\ \bibinfo {author} {\bibfnamefont {J.~M.}\ \bibnamefont
  {Doyle}},\ }\href@noop {} {\bibfield  {journal} {\bibinfo  {journal}
  {ChemPhysChem}\ }\textbf {\bibinfo {volume} {10}},\ \bibinfo {pages} {604}
  (\bibinfo {year} {2009})}\BibitemShut {NoStop}%
\bibitem [{\citenamefont {Capogrosso-Sansone}\ \emph
  {et~al.}(2010)\citenamefont {Capogrosso-Sansone}, \citenamefont {Trefzger},
  \citenamefont {Lewenstein}, \citenamefont {Zoller},\ and\ \citenamefont
  {Pupillo}}]{Capogrosso-Sansone2010}%
  \BibitemOpen
  \bibfield  {author} {\bibinfo {author} {\bibfnamefont {B.}~\bibnamefont
  {Capogrosso-Sansone}}, \bibinfo {author} {\bibfnamefont {C.}~\bibnamefont
  {Trefzger}}, \bibinfo {author} {\bibfnamefont {M.}~\bibnamefont
  {Lewenstein}}, \bibinfo {author} {\bibfnamefont {P.}~\bibnamefont {Zoller}},
  \ and\ \bibinfo {author} {\bibfnamefont {G.}~\bibnamefont {Pupillo}},\
  }\href@noop {} {\bibfield  {journal} {\bibinfo  {journal} {Phys. Rev. Lett.}\
  }\textbf {\bibinfo {volume} {104}},\ \bibinfo {pages} {125301} (\bibinfo
  {year} {2010})}\BibitemShut {NoStop}%
\bibitem [{\citenamefont {Micheli}\ \emph {et~al.}(2007)\citenamefont
  {Micheli}, \citenamefont {Pupillo}, \citenamefont {B\"uchler},\ and\
  \citenamefont {Zoller}}]{Micheli:2007}%
  \BibitemOpen
  \bibfield  {author} {\bibinfo {author} {\bibfnamefont {A.}~\bibnamefont
  {Micheli}}, \bibinfo {author} {\bibfnamefont {G.}~\bibnamefont {Pupillo}},
  \bibinfo {author} {\bibfnamefont {H.~P.}\ \bibnamefont {B\"uchler}}, \ and\
  \bibinfo {author} {\bibfnamefont {P.}~\bibnamefont {Zoller}},\ }\href@noop {}
  {\bibfield  {journal} {\bibinfo  {journal} {Phys. Rev. A}\ }\textbf {\bibinfo
  {volume} {76}},\ \bibinfo {pages} {043604} (\bibinfo {year}
  {2007})}\BibitemShut {NoStop}%
\bibitem [{\citenamefont {Wall}\ and\ \citenamefont {Carr}(2009)}]{Wall2009}%
  \BibitemOpen
  \bibfield  {author} {\bibinfo {author} {\bibfnamefont {M.~L.}\ \bibnamefont
  {Wall}}\ and\ \bibinfo {author} {\bibfnamefont {L.~D.}\ \bibnamefont
  {Carr}},\ }\href@noop {} {\bibfield  {journal} {\bibinfo  {journal} {New J.
  Phys.}\ }\textbf {\bibinfo {volume} {11}},\ \bibinfo {pages} {055027}
  (\bibinfo {year} {2009})}\BibitemShut {NoStop}%
\bibitem [{\citenamefont {Ni}\ \emph {et~al.}(2008)\citenamefont {Ni},
  \citenamefont {Ospelkaus}, \citenamefont {{de Miranda}}, \citenamefont
  {Pe'er}, \citenamefont {Neyenhuis}, \citenamefont {Zirbel}, \citenamefont
  {Kotochigova}, \citenamefont {Julienne}, \citenamefont {Jin},\ and\
  \citenamefont {Ye}}]{Ni:KRb:2008}%
  \BibitemOpen
  \bibfield  {author} {\bibinfo {author} {\bibfnamefont {K.-K.}\ \bibnamefont
  {Ni}}, \bibinfo {author} {\bibfnamefont {S.}~\bibnamefont {Ospelkaus}},
  \bibinfo {author} {\bibfnamefont {M.~H.~G.}\ \bibnamefont {{de Miranda}}},
  \bibinfo {author} {\bibfnamefont {A.}~\bibnamefont {Pe'er}}, \bibinfo
  {author} {\bibfnamefont {B.}~\bibnamefont {Neyenhuis}}, \bibinfo {author}
  {\bibfnamefont {J.~J.}\ \bibnamefont {Zirbel}}, \bibinfo {author}
  {\bibfnamefont {S.}~\bibnamefont {Kotochigova}}, \bibinfo {author}
  {\bibfnamefont {P.~S.}\ \bibnamefont {Julienne}}, \bibinfo {author}
  {\bibfnamefont {D.~S.}\ \bibnamefont {Jin}}, \ and\ \bibinfo {author}
  {\bibfnamefont {J.}~\bibnamefont {Ye}},\ }\href@noop {} {\bibfield  {journal}
  {\bibinfo  {journal} {Science}\ }\textbf {\bibinfo {volume} {322}},\ \bibinfo
  {pages} {231} (\bibinfo {year} {2008})}\BibitemShut {NoStop}%
\bibitem [{\citenamefont {Ospelkaus}\ \emph
  {et~al.}(2010{\natexlab{a}})\citenamefont {Ospelkaus}, \citenamefont {Ni},
  \citenamefont {Qu\'{e}m\'{e}ner}, \citenamefont {Neyenhuis}, \citenamefont
  {Wang}, \citenamefont {{de Miranda}}, \citenamefont {Bohn}, \citenamefont
  {Ye},\ and\ \citenamefont {Jin}}]{Ospelkaus:hyperfine-control:2010}%
  \BibitemOpen
  \bibfield  {author} {\bibinfo {author} {\bibfnamefont {S.}~\bibnamefont
  {Ospelkaus}}, \bibinfo {author} {\bibfnamefont {K.-K.}\ \bibnamefont {Ni}},
  \bibinfo {author} {\bibfnamefont {G.}~\bibnamefont {Qu\'{e}m\'{e}ner}},
  \bibinfo {author} {\bibfnamefont {B.}~\bibnamefont {Neyenhuis}}, \bibinfo
  {author} {\bibfnamefont {D.}~\bibnamefont {Wang}}, \bibinfo {author}
  {\bibfnamefont {M.~H.~G.}\ \bibnamefont {{de Miranda}}}, \bibinfo {author}
  {\bibfnamefont {J.~L.}\ \bibnamefont {Bohn}}, \bibinfo {author}
  {\bibfnamefont {J.}~\bibnamefont {Ye}}, \ and\ \bibinfo {author}
  {\bibfnamefont {D.~S.}\ \bibnamefont {Jin}},\ }\href@noop {} {\bibfield
  {journal} {\bibinfo  {journal} {Phys. Rev. Lett.}\ }\textbf {\bibinfo
  {volume} {104}},\ \bibinfo {pages} {030402} (\bibinfo {year}
  {2010}{\natexlab{a}})}\BibitemShut {NoStop}%
\bibitem [{\citenamefont {de~Miranda}\ \emph {et~al.}(2011)\citenamefont
  {de~Miranda}, \citenamefont {Chotia}, \citenamefont {Neyenhuis},
  \citenamefont {Wang}, \citenamefont {Qu\'em\'ener}, \citenamefont
  {Ospelkaus}, \citenamefont {Bohn}, \citenamefont {Ye},\ and\ \citenamefont
  {Jin}}]{deMiranda:2011}%
  \BibitemOpen
  \bibfield  {author} {\bibinfo {author} {\bibfnamefont {M.~H.~G.}\
  \bibnamefont {de~Miranda}}, \bibinfo {author} {\bibfnamefont
  {A.}~\bibnamefont {Chotia}}, \bibinfo {author} {\bibfnamefont
  {B.}~\bibnamefont {Neyenhuis}}, \bibinfo {author} {\bibfnamefont
  {D.}~\bibnamefont {Wang}}, \bibinfo {author} {\bibfnamefont {G.}~\bibnamefont
  {Qu\'em\'ener}}, \bibinfo {author} {\bibfnamefont {S.}~\bibnamefont
  {Ospelkaus}}, \bibinfo {author} {\bibfnamefont {J.~L.}\ \bibnamefont {Bohn}},
  \bibinfo {author} {\bibfnamefont {J.}~\bibnamefont {Ye}}, \ and\ \bibinfo
  {author} {\bibfnamefont {D.~S.}\ \bibnamefont {Jin}},\ }\href@noop {}
  {\bibfield  {journal} {\bibinfo  {journal} {Nat. Phys.}\ }\textbf {\bibinfo
  {volume} {7}},\ \bibinfo {pages} {502} (\bibinfo {year} {2011})}\BibitemShut
  {NoStop}%
\bibitem [{\citenamefont {Chotia}\ \emph {et~al.}(2012)\citenamefont {Chotia},
  \citenamefont {Neyenhuis}, \citenamefont {Moses}, \citenamefont {Yan},
  \citenamefont {Covey}, \citenamefont {Foss-Feig}, \citenamefont {Rey},
  \citenamefont {Jin},\ and\ \citenamefont {Ye}}]{Chotia:2012}%
  \BibitemOpen
  \bibfield  {author} {\bibinfo {author} {\bibfnamefont {A.}~\bibnamefont
  {Chotia}}, \bibinfo {author} {\bibfnamefont {B.}~\bibnamefont {Neyenhuis}},
  \bibinfo {author} {\bibfnamefont {S.~A.}\ \bibnamefont {Moses}}, \bibinfo
  {author} {\bibfnamefont {B.}~\bibnamefont {Yan}}, \bibinfo {author}
  {\bibfnamefont {J.~P.}\ \bibnamefont {Covey}}, \bibinfo {author}
  {\bibfnamefont {M.}~\bibnamefont {Foss-Feig}}, \bibinfo {author}
  {\bibfnamefont {A.~M.}\ \bibnamefont {Rey}}, \bibinfo {author} {\bibfnamefont
  {D.~S.}\ \bibnamefont {Jin}}, \ and\ \bibinfo {author} {\bibfnamefont
  {J.}~\bibnamefont {Ye}},\ }\href@noop {} {\bibfield  {journal} {\bibinfo
  {journal} {Phys. rev. Lett.}\ }\textbf {\bibinfo {volume} {108}},\ \bibinfo
  {pages} {080405} (\bibinfo {year} {2012})}\BibitemShut {NoStop}%
\bibitem [{\citenamefont {Danzl}\ \emph {et~al.}(2010)\citenamefont {Danzl},
  \citenamefont {Mark}, \citenamefont {Haller}, \citenamefont {Gustavsson},
  \citenamefont {Hart}, \citenamefont {Aldegunde}, \citenamefont {Hutson},\
  and\ \citenamefont {N\"agerl}}]{Danzl:ground:2010}%
  \BibitemOpen
  \bibfield  {author} {\bibinfo {author} {\bibfnamefont {J.~G.}\ \bibnamefont
  {Danzl}}, \bibinfo {author} {\bibfnamefont {M.~J.}\ \bibnamefont {Mark}},
  \bibinfo {author} {\bibfnamefont {E.}~\bibnamefont {Haller}}, \bibinfo
  {author} {\bibfnamefont {M.}~\bibnamefont {Gustavsson}}, \bibinfo {author}
  {\bibfnamefont {R.}~\bibnamefont {Hart}}, \bibinfo {author} {\bibfnamefont
  {J.}~\bibnamefont {Aldegunde}}, \bibinfo {author} {\bibfnamefont {J.~M.}\
  \bibnamefont {Hutson}}, \ and\ \bibinfo {author} {\bibfnamefont {H.-C.}\
  \bibnamefont {N\"agerl}},\ }\href {\doibase doi:10.1038/nphys1533} {\bibfield
   {journal} {\bibinfo  {journal} {Nature Phys.}\ }\textbf {\bibinfo {volume}
  {6}},\ \bibinfo {pages} {265} (\bibinfo {year} {2010})}\BibitemShut {NoStop}%
\bibitem [{\citenamefont {Lang}\ \emph {et~al.}(2008)\citenamefont {Lang},
  \citenamefont {Winkler}, \citenamefont {Strauss}, \citenamefont {Grimm},\
  and\ \citenamefont {Hecker~Denschlag}}]{Lang:ground:2008}%
  \BibitemOpen
  \bibfield  {author} {\bibinfo {author} {\bibfnamefont {F.}~\bibnamefont
  {Lang}}, \bibinfo {author} {\bibfnamefont {K.}~\bibnamefont {Winkler}},
  \bibinfo {author} {\bibfnamefont {C.}~\bibnamefont {Strauss}}, \bibinfo
  {author} {\bibfnamefont {R.}~\bibnamefont {Grimm}}, \ and\ \bibinfo {author}
  {\bibfnamefont {J.}~\bibnamefont {Hecker~Denschlag}},\ }\href@noop {}
  {\bibfield  {journal} {\bibinfo  {journal} {Phys. Rev. Lett.}\ }\textbf
  {\bibinfo {volume} {101}},\ \bibinfo {pages} {133005} (\bibinfo {year}
  {2008})}\BibitemShut {NoStop}%
\bibitem [{\citenamefont {Ospelkaus}\ \emph
  {et~al.}(2010{\natexlab{b}})\citenamefont {Ospelkaus}, \citenamefont {Ni},
  \citenamefont {Wang}, \citenamefont {{de Miranda}}, \citenamefont
  {Neyenhuis}, \citenamefont {Qu\'{e}m\'{e}ner}, \citenamefont {Julienne},
  \citenamefont {Bohn}, \citenamefont {Jin},\ and\ \citenamefont
  {Ye}}]{Ospelkaus:react:2010}%
  \BibitemOpen
  \bibfield  {author} {\bibinfo {author} {\bibfnamefont {S.}~\bibnamefont
  {Ospelkaus}}, \bibinfo {author} {\bibfnamefont {K.-K.}\ \bibnamefont {Ni}},
  \bibinfo {author} {\bibfnamefont {D.}~\bibnamefont {Wang}}, \bibinfo {author}
  {\bibfnamefont {M.~H.~G.}\ \bibnamefont {{de Miranda}}}, \bibinfo {author}
  {\bibfnamefont {B.}~\bibnamefont {Neyenhuis}}, \bibinfo {author}
  {\bibfnamefont {G.}~\bibnamefont {Qu\'{e}m\'{e}ner}}, \bibinfo {author}
  {\bibfnamefont {P.~S.}\ \bibnamefont {Julienne}}, \bibinfo {author}
  {\bibfnamefont {J.~L.}\ \bibnamefont {Bohn}}, \bibinfo {author}
  {\bibfnamefont {D.~S.}\ \bibnamefont {Jin}}, \ and\ \bibinfo {author}
  {\bibfnamefont {J.}~\bibnamefont {Ye}},\ }\href@noop {} {\bibfield  {journal}
  {\bibinfo  {journal} {Science}\ }\textbf {\bibinfo {volume} {327}},\ \bibinfo
  {pages} {853} (\bibinfo {year} {2010}{\natexlab{b}})}\BibitemShut {NoStop}%
\bibitem [{\citenamefont {\.Zuchowski}\ and\ \citenamefont
  {Hutson}(2010)}]{Zuchowski:trimers:2010}%
  \BibitemOpen
  \bibfield  {author} {\bibinfo {author} {\bibfnamefont {P.~S.}\ \bibnamefont
  {\.Zuchowski}}\ and\ \bibinfo {author} {\bibfnamefont {J.~M.}\ \bibnamefont
  {Hutson}},\ }\href@noop {} {\bibfield  {journal} {\bibinfo  {journal} {Phys.
  Rev. A}\ }\textbf {\bibinfo {volume} {81}},\ \bibinfo {pages} {060703(R)}
  (\bibinfo {year} {2010})}\BibitemShut {NoStop}%
\bibitem [{\citenamefont {Wu}\ \emph {et~al.}(2012)\citenamefont {Wu},
  \citenamefont {Park}, \citenamefont {Ahmadi}, \citenamefont {Will},\ and\
  \citenamefont {Zwierlein}}]{Wu:2012}%
  \BibitemOpen
  \bibfield  {author} {\bibinfo {author} {\bibfnamefont {C.-H.}\ \bibnamefont
  {Wu}}, \bibinfo {author} {\bibfnamefont {J.~W.}\ \bibnamefont {Park}},
  \bibinfo {author} {\bibfnamefont {P.}~\bibnamefont {Ahmadi}}, \bibinfo
  {author} {\bibfnamefont {S.}~\bibnamefont {Will}}, \ and\ \bibinfo {author}
  {\bibfnamefont {M.~W.}\ \bibnamefont {Zwierlein}},\ }\href@noop {} {\bibfield
   {journal} {\bibinfo  {journal} {Phys. Rev. Lett.}\ }\textbf {\bibinfo
  {volume} {109}},\ \bibinfo {pages} {085301} (\bibinfo {year}
  {2012})}\BibitemShut {NoStop}%
\bibitem [{\citenamefont {Takekoshi}\ \emph {et~al.}(2012)\citenamefont
  {Takekoshi}, \citenamefont {Debatin}, \citenamefont {Rameshan}, \citenamefont
  {Ferlaino}, \citenamefont {Grimm}, \citenamefont {N\"agerl}, \citenamefont
  {{Le Sueur}}, \citenamefont {Hutson}, \citenamefont {Julienne}, \citenamefont
  {Kotochigova},\ and\ \citenamefont {Tiemann}}]{Takekoshi:RbCs:2012}%
  \BibitemOpen
  \bibfield  {author} {\bibinfo {author} {\bibfnamefont {T.}~\bibnamefont
  {Takekoshi}}, \bibinfo {author} {\bibfnamefont {M.}~\bibnamefont {Debatin}},
  \bibinfo {author} {\bibfnamefont {R.}~\bibnamefont {Rameshan}}, \bibinfo
  {author} {\bibfnamefont {F.}~\bibnamefont {Ferlaino}}, \bibinfo {author}
  {\bibfnamefont {R.}~\bibnamefont {Grimm}}, \bibinfo {author} {\bibfnamefont
  {H.-C.}\ \bibnamefont {N\"agerl}}, \bibinfo {author} {\bibfnamefont {C.~R.}\
  \bibnamefont {{Le Sueur}}}, \bibinfo {author} {\bibfnamefont {J.~M.}\
  \bibnamefont {Hutson}}, \bibinfo {author} {\bibfnamefont {P.~S.}\
  \bibnamefont {Julienne}}, \bibinfo {author} {\bibfnamefont {S.}~\bibnamefont
  {Kotochigova}}, \ and\ \bibinfo {author} {\bibfnamefont {E.}~\bibnamefont
  {Tiemann}},\ }\href@noop {} {\bibfield  {journal} {\bibinfo  {journal} {Phys.
  Rev. A}\ }\textbf {\bibinfo {volume} {85}},\ \bibinfo {pages} {032506}
  (\bibinfo {year} {2012})}\BibitemShut {NoStop}%
\bibitem [{\citenamefont {K\"oppinger}\ \emph {et~al.}(2014)\citenamefont
  {K\"oppinger}, , \citenamefont {McCarron}, \citenamefont {Jenkin},
  \citenamefont {Molony}, \citenamefont {Cho}, \citenamefont {Cornish},
  \citenamefont {{Le Sueur}}, \citenamefont {Blackley},\ and\ \citenamefont
  {Hutson}}]{Koppinger:RbCs:2014}%
  \BibitemOpen
  \bibfield  {author} {\bibinfo {author} {\bibfnamefont {M.~P.}\ \bibnamefont
  {K\"oppinger}}, , \bibinfo {author} {\bibfnamefont {D.~J.}\ \bibnamefont
  {McCarron}}, \bibinfo {author} {\bibfnamefont {D.~L.}\ \bibnamefont
  {Jenkin}}, \bibinfo {author} {\bibfnamefont {P.~K.}\ \bibnamefont {Molony}},
  \bibinfo {author} {\bibfnamefont {H.-W.}\ \bibnamefont {Cho}}, \bibinfo
  {author} {\bibfnamefont {S.~L.}\ \bibnamefont {Cornish}}, \bibinfo {author}
  {\bibfnamefont {C.~R.}\ \bibnamefont {{Le Sueur}}}, \bibinfo {author}
  {\bibfnamefont {C.~L.}\ \bibnamefont {Blackley}}, \ and\ \bibinfo {author}
  {\bibfnamefont {J.~M.}\ \bibnamefont {Hutson}},\ }\href@noop {} {\bibfield
  {journal} {\bibinfo  {journal} {Phys. Rev. A}\ }\textbf {\bibinfo {volume}
  {89}},\ \bibinfo {pages} {033604} (\bibinfo {year} {2014})}\BibitemShut
  {NoStop}%
\bibitem [{\citenamefont {Takekoshi}\ \emph {et~al.}(2014)\citenamefont
  {Takekoshi}, \citenamefont {Reichs\"ollner}, \citenamefont {Schindewolf},
  \citenamefont {Hutson}, \citenamefont {{Le Sueur}}, \citenamefont {Dulieu},
  \citenamefont {Ferlaino}, \citenamefont {Grimm},\ and\ \citenamefont
  {N\"agerl}}]{Takekoshi:RbCs:2014}%
  \BibitemOpen
  \bibfield  {author} {\bibinfo {author} {\bibfnamefont {T.}~\bibnamefont
  {Takekoshi}}, \bibinfo {author} {\bibfnamefont {L.}~\bibnamefont
  {Reichs\"ollner}}, \bibinfo {author} {\bibfnamefont {A.}~\bibnamefont
  {Schindewolf}}, \bibinfo {author} {\bibfnamefont {J.~M.}\ \bibnamefont
  {Hutson}}, \bibinfo {author} {\bibfnamefont {C.~R.}\ \bibnamefont {{Le
  Sueur}}}, \bibinfo {author} {\bibfnamefont {O.}~\bibnamefont {Dulieu}},
  \bibinfo {author} {\bibfnamefont {F.}~\bibnamefont {Ferlaino}}, \bibinfo
  {author} {\bibfnamefont {R.}~\bibnamefont {Grimm}}, \ and\ \bibinfo {author}
  {\bibfnamefont {H.-C.}\ \bibnamefont {N\"agerl}},\ }\href@noop {} {\bibfield
  {journal} {\bibinfo  {journal} {arXiv:1405.6037}\ } (\bibinfo {year}
  {2014})}\BibitemShut {NoStop}%
\bibitem [{\citenamefont {Aymar}\ and\ \citenamefont
  {Dulieu}(2005)}]{Aymar:2005}%
  \BibitemOpen
  \bibfield  {author} {\bibinfo {author} {\bibfnamefont {M.}~\bibnamefont
  {Aymar}}\ and\ \bibinfo {author} {\bibfnamefont {O.}~\bibnamefont {Dulieu}},\
  }\href@noop {} {\bibfield  {journal} {\bibinfo  {journal} {J. Chem. Phys.}\
  }\textbf {\bibinfo {volume} {122}},\ \bibinfo {pages} {204302} (\bibinfo
  {year} {2005})}\BibitemShut {NoStop}%
\bibitem [{\citenamefont {Julienne}\ \emph {et~al.}(2011)\citenamefont
  {Julienne}, \citenamefont {Hanna},\ and\ \citenamefont
  {Idziaszek}}]{Julienne:dipoles:2011}%
  \BibitemOpen
  \bibfield  {author} {\bibinfo {author} {\bibfnamefont {P.~S.}\ \bibnamefont
  {Julienne}}, \bibinfo {author} {\bibfnamefont {T.~M.}\ \bibnamefont {Hanna}},
  \ and\ \bibinfo {author} {\bibfnamefont {Z.}~\bibnamefont {Idziaszek}},\
  }\href@noop {} {\bibfield  {journal} {\bibinfo  {journal} {Phys. Chem. Chem.
  Phys.}\ }\textbf {\bibinfo {volume} {13}},\ \bibinfo {pages} {19114}
  (\bibinfo {year} {2011})}\BibitemShut {NoStop}%
\bibitem [{\citenamefont {Ferber}\ \emph {et~al.}(2009)\citenamefont {Ferber},
  \citenamefont {Klincare}, \citenamefont {Nikolayeva}, \citenamefont
  {Tamanis}, \citenamefont {Kn\"ockel}, \citenamefont {Tiemann},\ and\
  \citenamefont {Pashov}}]{Ferber:2009}%
  \BibitemOpen
  \bibfield  {author} {\bibinfo {author} {\bibfnamefont {R.}~\bibnamefont
  {Ferber}}, \bibinfo {author} {\bibfnamefont {I.}~\bibnamefont {Klincare}},
  \bibinfo {author} {\bibfnamefont {O.}~\bibnamefont {Nikolayeva}}, \bibinfo
  {author} {\bibfnamefont {M.}~\bibnamefont {Tamanis}}, \bibinfo {author}
  {\bibfnamefont {H.}~\bibnamefont {Kn\"ockel}}, \bibinfo {author}
  {\bibfnamefont {E.}~\bibnamefont {Tiemann}}, \ and\ \bibinfo {author}
  {\bibfnamefont {A.}~\bibnamefont {Pashov}},\ }\href@noop {} {\bibfield
  {journal} {\bibinfo  {journal} {Phys. Rev. A}\ }\textbf {\bibinfo {volume}
  {80}},\ \bibinfo {pages} {062501} (\bibinfo {year} {2009})}\BibitemShut
  {NoStop}%
\bibitem [{\citenamefont {Ferber}\ \emph {et~al.}(2008)\citenamefont {Ferber},
  \citenamefont {Klincare}, \citenamefont {Nikolayeva}, \citenamefont
  {Tamanis}, \citenamefont {Kn\"ockel}, \citenamefont {Tiemann},\ and\
  \citenamefont {Pashov}}]{Ferber:2008}%
  \BibitemOpen
  \bibfield  {author} {\bibinfo {author} {\bibfnamefont {R.}~\bibnamefont
  {Ferber}}, \bibinfo {author} {\bibfnamefont {I.}~\bibnamefont {Klincare}},
  \bibinfo {author} {\bibfnamefont {O.}~\bibnamefont {Nikolayeva}}, \bibinfo
  {author} {\bibfnamefont {M.}~\bibnamefont {Tamanis}}, \bibinfo {author}
  {\bibfnamefont {H.}~\bibnamefont {Kn\"ockel}}, \bibinfo {author}
  {\bibfnamefont {E.}~\bibnamefont {Tiemann}}, \ and\ \bibinfo {author}
  {\bibfnamefont {A.}~\bibnamefont {Pashov}},\ }\href@noop {} {\bibfield
  {journal} {\bibinfo  {journal} {J. Chem. Phys.}\ }\textbf {\bibinfo {volume}
  {128}},\ \bibinfo {pages} {244316} (\bibinfo {year} {2008})}\BibitemShut
  {NoStop}%
\bibitem [{\citenamefont {Ferber}\ \emph {et~al.}(2013)\citenamefont {Ferber},
  \citenamefont {Nikolayeva}, \citenamefont {Tamanis}, \citenamefont
  {Kn\"ockel},\ and\ \citenamefont {Tiemann}}]{Ferber:2013}%
  \BibitemOpen
  \bibfield  {author} {\bibinfo {author} {\bibfnamefont {R.}~\bibnamefont
  {Ferber}}, \bibinfo {author} {\bibfnamefont {O.}~\bibnamefont {Nikolayeva}},
  \bibinfo {author} {\bibfnamefont {M.}~\bibnamefont {Tamanis}}, \bibinfo
  {author} {\bibfnamefont {H.}~\bibnamefont {Kn\"ockel}}, \ and\ \bibinfo
  {author} {\bibfnamefont {E.}~\bibnamefont {Tiemann}},\ }\href@noop {}
  {\bibfield  {journal} {\bibinfo  {journal} {Phys. Rev. A}\ }\textbf {\bibinfo
  {volume} {88}},\ \bibinfo {pages} {012516} (\bibinfo {year}
  {2013})}\BibitemShut {NoStop}%
\bibitem [{\citenamefont {Hutson}\ and\ \citenamefont
  {Green}(1994)}]{molscat:v14}%
  \BibitemOpen
  \bibfield  {author} {\bibinfo {author} {\bibfnamefont {J.~M.}\ \bibnamefont
  {Hutson}}\ and\ \bibinfo {author} {\bibfnamefont {S.}~\bibnamefont {Green}},\
  }\href@noop {} {\enquote {\bibinfo {title} {{MOLSCAT} computer program,
  version 14},}\ }\bibinfo {howpublished} {distributed by Collaborative
  Computational Project No.\ 6 of the UK Engineering and Physical Sciences
  Research Council} (\bibinfo {year} {1994})\BibitemShut {NoStop}%
\bibitem [{\citenamefont {Hutson}(1993)}]{Hutson:bound:1993}%
  \BibitemOpen
  \bibfield  {author} {\bibinfo {author} {\bibfnamefont {J.~M.}\ \bibnamefont
  {Hutson}},\ }\href@noop {} {\enquote {\bibinfo {title} {{BOUND} computer
  program, version 5},}\ }\bibinfo {howpublished} {distributed by Collaborative
  Computational Project No.\ 6 of the UK Engineering and Physical Sciences
  Research Council} (\bibinfo {year} {1993})\BibitemShut {NoStop}%
\bibitem [{\citenamefont {Gonz\'{a}lez-Mart\'{\i}nez}\ and\ \citenamefont
  {Hutson}(2007)}]{Gonzalez-Martinez:2007}%
  \BibitemOpen
  \bibfield  {author} {\bibinfo {author} {\bibfnamefont {M.~L.}\ \bibnamefont
  {Gonz\'{a}lez-Mart\'{\i}nez}}\ and\ \bibinfo {author} {\bibfnamefont {J.~M.}\
  \bibnamefont {Hutson}},\ }\href@noop {} {\bibfield  {journal} {\bibinfo
  {journal} {Phys. Rev. A}\ }\textbf {\bibinfo {volume} {75}},\ \bibinfo
  {pages} {022702} (\bibinfo {year} {2007})}\BibitemShut {NoStop}%
\bibitem [{\citenamefont {Berninger}\ \emph {et~al.}(2013)\citenamefont
  {Berninger}, \citenamefont {Zenesini}, \citenamefont {Huang}, \citenamefont
  {Harm}, \citenamefont {N\"agerl}, \citenamefont {Ferlaino}, \citenamefont
  {Grimm}, \citenamefont {Julienne},\ and\ \citenamefont
  {Hutson}}]{Berninger:Cs2:2013}%
  \BibitemOpen
  \bibfield  {author} {\bibinfo {author} {\bibfnamefont {M.}~\bibnamefont
  {Berninger}}, \bibinfo {author} {\bibfnamefont {A.}~\bibnamefont {Zenesini}},
  \bibinfo {author} {\bibfnamefont {B.}~\bibnamefont {Huang}}, \bibinfo
  {author} {\bibfnamefont {W.}~\bibnamefont {Harm}}, \bibinfo {author}
  {\bibfnamefont {H.-C.}\ \bibnamefont {N\"agerl}}, \bibinfo {author}
  {\bibfnamefont {F.}~\bibnamefont {Ferlaino}}, \bibinfo {author}
  {\bibfnamefont {R.}~\bibnamefont {Grimm}}, \bibinfo {author} {\bibfnamefont
  {P.~S.}\ \bibnamefont {Julienne}}, \ and\ \bibinfo {author} {\bibfnamefont
  {J.~M.}\ \bibnamefont {Hutson}},\ }\href@noop {} {\bibfield  {journal}
  {\bibinfo  {journal} {Phys. Rev. A}\ }\textbf {\bibinfo {volume} {87}},\
  \bibinfo {pages} {032517} (\bibinfo {year} {2013})}\BibitemShut {NoStop}%
\bibitem [{\citenamefont {Hutson}\ \emph {et~al.}(2008)\citenamefont {Hutson},
  \citenamefont {Tiesinga},\ and\ \citenamefont
  {Julienne}}]{Hutson:Cs2-note:2008}%
  \BibitemOpen
  \bibfield  {author} {\bibinfo {author} {\bibfnamefont {J.~M.}\ \bibnamefont
  {Hutson}}, \bibinfo {author} {\bibfnamefont {E.}~\bibnamefont {Tiesinga}}, \
  and\ \bibinfo {author} {\bibfnamefont {P.~S.}\ \bibnamefont {Julienne}},\
  }\href@noop {} {\bibfield  {journal} {\bibinfo  {journal} {Phys. Rev. A}\
  }\textbf {\bibinfo {volume} {78}},\ \bibinfo {pages} {052703} (\bibinfo
  {year} {2008})},\ \bibinfo {note} {note that the matrix element of the
  dipolar spin-spin operator given in Eq.\ A2 of this paper omits a factor of
  $-\sqrt{30}$.}\BibitemShut {Stop}%
\bibitem [{\citenamefont {Hutson}(2007)}]{Hutson:res:2007}%
  \BibitemOpen
  \bibfield  {author} {\bibinfo {author} {\bibfnamefont {J.~M.}\ \bibnamefont
  {Hutson}},\ }\href@noop {} {\bibfield  {journal} {\bibinfo  {journal} {New J.
  Phys.}\ }\textbf {\bibinfo {volume} {9}},\ \bibinfo {pages} {152} (\bibinfo
  {year} {2007})}\BibitemShut {NoStop}%
\bibitem [{\citenamefont {Hutson}(2011)}]{Hutson:field:2011}%
  \BibitemOpen
  \bibfield  {author} {\bibinfo {author} {\bibfnamefont {J.~M.}\ \bibnamefont
  {Hutson}},\ }\href@noop {} {\enquote {\bibinfo {title} {{FIELD} computer
  program, version 1},}\ } (\bibinfo {year} {2011})\BibitemShut {NoStop}%
\bibitem [{\citenamefont {Moerdijk}\ \emph {et~al.}(1995)\citenamefont
  {Moerdijk}, \citenamefont {Verhaar},\ and\ \citenamefont
  {Axelsson}}]{Moerdijk:1995}%
  \BibitemOpen
  \bibfield  {author} {\bibinfo {author} {\bibfnamefont {A.~J.}\ \bibnamefont
  {Moerdijk}}, \bibinfo {author} {\bibfnamefont {B.~J.}\ \bibnamefont
  {Verhaar}}, \ and\ \bibinfo {author} {\bibfnamefont {A.}~\bibnamefont
  {Axelsson}},\ }\href@noop {} {\bibfield  {journal} {\bibinfo  {journal}
  {Phys. Rev. A}\ }\textbf {\bibinfo {volume} {51}},\ \bibinfo {pages} {4852}
  (\bibinfo {year} {1995})}\BibitemShut {NoStop}%
\bibitem [{\citenamefont {Falke}\ \emph {et~al.}(2008)\citenamefont {Falke},
  \citenamefont {Kn\"ockel}, \citenamefont {Friebe}, \citenamefont {Riedmann},
  \citenamefont {Tiemann},\ and\ \citenamefont {Lisdat}}]{Falke:2008}%
  \BibitemOpen
  \bibfield  {author} {\bibinfo {author} {\bibfnamefont {S.}~\bibnamefont
  {Falke}}, \bibinfo {author} {\bibfnamefont {H.}~\bibnamefont {Kn\"ockel}},
  \bibinfo {author} {\bibfnamefont {J.}~\bibnamefont {Friebe}}, \bibinfo
  {author} {\bibfnamefont {M.}~\bibnamefont {Riedmann}}, \bibinfo {author}
  {\bibfnamefont {E.}~\bibnamefont {Tiemann}}, \ and\ \bibinfo {author}
  {\bibfnamefont {C.}~\bibnamefont {Lisdat}},\ }\href@noop {} {\bibfield
  {journal} {\bibinfo  {journal} {Phys. Rev. A}\ }\textbf {\bibinfo {volume}
  {78}},\ \bibinfo {pages} {012503} (\bibinfo {year} {2008})}\BibitemShut
  {NoStop}%
\bibitem [{\citenamefont {Stoof}\ \emph {et~al.}(1988)\citenamefont {Stoof},
  \citenamefont {Koelman},\ and\ \citenamefont {Verhaar}}]{Stoof:1988}%
  \BibitemOpen
  \bibfield  {author} {\bibinfo {author} {\bibfnamefont {H.~T.~C.}\
  \bibnamefont {Stoof}}, \bibinfo {author} {\bibfnamefont {J.~M. V.~A.}\
  \bibnamefont {Koelman}}, \ and\ \bibinfo {author} {\bibfnamefont {B.~J.}\
  \bibnamefont {Verhaar}},\ }\href@noop {} {\bibfield  {journal} {\bibinfo
  {journal} {Phys. Rev. B}\ }\textbf {\bibinfo {volume} {38}},\ \bibinfo
  {pages} {4688} (\bibinfo {year} {1988})}\BibitemShut {NoStop}%
\bibitem [{\citenamefont {Mies}\ \emph {et~al.}(1996)\citenamefont {Mies},
  \citenamefont {Williams}, \citenamefont {Julienne},\ and\ \citenamefont
  {Krauss}}]{Mies:1996}%
  \BibitemOpen
  \bibfield  {author} {\bibinfo {author} {\bibfnamefont {F.~H.}\ \bibnamefont
  {Mies}}, \bibinfo {author} {\bibfnamefont {C.~J.}\ \bibnamefont {Williams}},
  \bibinfo {author} {\bibfnamefont {P.~S.}\ \bibnamefont {Julienne}}, \ and\
  \bibinfo {author} {\bibfnamefont {M.}~\bibnamefont {Krauss}},\ }\href@noop {}
  {\bibfield  {journal} {\bibinfo  {journal} {J. Res. Natl. Inst. Stand.
  Technol.}\ }\textbf {\bibinfo {volume} {101}},\ \bibinfo {pages} {521}
  (\bibinfo {year} {1996})}\BibitemShut {NoStop}%
\bibitem [{\citenamefont {Kotochigova}\ \emph {et~al.}(2000)\citenamefont
  {Kotochigova}, \citenamefont {Tiesinga},\ and\ \citenamefont
  {Julienne}}]{Kotochigova:2001}%
  \BibitemOpen
  \bibfield  {author} {\bibinfo {author} {\bibfnamefont {S.}~\bibnamefont
  {Kotochigova}}, \bibinfo {author} {\bibfnamefont {E.}~\bibnamefont
  {Tiesinga}}, \ and\ \bibinfo {author} {\bibfnamefont {P.~S.}\ \bibnamefont
  {Julienne}},\ }\href@noop {} {\bibfield  {journal} {\bibinfo  {journal}
  {Phys. Rev. A}\ }\textbf {\bibinfo {volume} {63}},\ \bibinfo {pages} {012517}
  (\bibinfo {year} {2000})}\BibitemShut {NoStop}%
\bibitem [{Note1()}]{Note1}%
  \BibitemOpen
  \bibinfo {note} {See Supplemental Material at [to be inserted by publisher]
  for a full listing of the resonance positions and widths.}\BibitemShut
  {Stop}%
\bibitem [{\citenamefont {Julienne}\ \emph {et~al.}(1997)\citenamefont
  {Julienne}, \citenamefont {Mies}, \citenamefont {Tiesinga},\ and\
  \citenamefont {Williams}}]{Julienne:1997}%
  \BibitemOpen
  \bibfield  {author} {\bibinfo {author} {\bibfnamefont {P.~S.}\ \bibnamefont
  {Julienne}}, \bibinfo {author} {\bibfnamefont {F.~H.}\ \bibnamefont {Mies}},
  \bibinfo {author} {\bibfnamefont {E.}~\bibnamefont {Tiesinga}}, \ and\
  \bibinfo {author} {\bibfnamefont {C.~J.}\ \bibnamefont {Williams}},\
  }\href@noop {} {\bibfield  {journal} {\bibinfo  {journal} {Phys. Rev. Lett.}\
  }\textbf {\bibinfo {volume} {78}},\ \bibinfo {pages} {1880} (\bibinfo {year}
  {1997})}\BibitemShut {NoStop}%
\bibitem [{\citenamefont {Chin}\ \emph {et~al.}(2010)\citenamefont {Chin},
  \citenamefont {Grimm}, \citenamefont {Julienne},\ and\ \citenamefont
  {Tiesinga}}]{ChinRMP2010}%
  \BibitemOpen
  \bibfield  {author} {\bibinfo {author} {\bibfnamefont {C.}~\bibnamefont
  {Chin}}, \bibinfo {author} {\bibfnamefont {R.}~\bibnamefont {Grimm}},
  \bibinfo {author} {\bibfnamefont {P.}~\bibnamefont {Julienne}}, \ and\
  \bibinfo {author} {\bibfnamefont {E.}~\bibnamefont {Tiesinga}},\ }\href
  {\doibase 10.1103/RevModPhys.82.1225} {\bibfield  {journal} {\bibinfo
  {journal} {Rev. Mod. Phys.}\ }\textbf {\bibinfo {volume} {82}},\ \bibinfo
  {pages} {1225} (\bibinfo {year} {2010})}\BibitemShut {NoStop}%
\bibitem [{\citenamefont {Hodby}\ \emph {et~al.}(2005)\citenamefont {Hodby},
  \citenamefont {Thompson}, \citenamefont {Regal}, \citenamefont {Greiner},
  \citenamefont {Wilson}, \citenamefont {Jin}, \citenamefont {Cornell},\ and\
  \citenamefont {Wieman}}]{HodbyPRL2005}%
  \BibitemOpen
  \bibfield  {author} {\bibinfo {author} {\bibfnamefont {E.}~\bibnamefont
  {Hodby}}, \bibinfo {author} {\bibfnamefont {S.~T.}\ \bibnamefont {Thompson}},
  \bibinfo {author} {\bibfnamefont {C.~A.}\ \bibnamefont {Regal}}, \bibinfo
  {author} {\bibfnamefont {M.}~\bibnamefont {Greiner}}, \bibinfo {author}
  {\bibfnamefont {A.~C.}\ \bibnamefont {Wilson}}, \bibinfo {author}
  {\bibfnamefont {D.~S.}\ \bibnamefont {Jin}}, \bibinfo {author} {\bibfnamefont
  {E.~A.}\ \bibnamefont {Cornell}}, \ and\ \bibinfo {author} {\bibfnamefont
  {C.~E.}\ \bibnamefont {Wieman}},\ }\href {\doibase
  10.1103/PhysRevLett.94.120402} {\bibfield  {journal} {\bibinfo  {journal}
  {Phys. Rev. Lett.}\ }\textbf {\bibinfo {volume} {94}},\ \bibinfo {pages}
  {120402} (\bibinfo {year} {2005})}\BibitemShut {NoStop}%
\bibitem [{\citenamefont {Ketterle}\ and\ \citenamefont {van
  Druten}(1996)}]{KetterleAdv1996}%
  \BibitemOpen
  \bibfield  {author} {\bibinfo {author} {\bibfnamefont {W.}~\bibnamefont
  {Ketterle}}\ and\ \bibinfo {author} {\bibfnamefont {N.}~\bibnamefont {van
  Druten}},\ }\href@noop {} {\bibfield  {journal} {\bibinfo  {journal} {Adv.
  At. Mol. Opt. Phys.}\ }\textbf {\bibinfo {volume} {37}},\ \bibinfo {pages}
  {181} (\bibinfo {year} {1996})}\BibitemShut {NoStop}%
\bibitem [{\citenamefont {Weiner}\ \emph {et~al.}(1999)\citenamefont {Weiner},
  \citenamefont {Bagnato}, \citenamefont {Zilio},\ and\ \citenamefont
  {Julienne}}]{WeinerRMP1999}%
  \BibitemOpen
  \bibfield  {author} {\bibinfo {author} {\bibfnamefont {J.}~\bibnamefont
  {Weiner}}, \bibinfo {author} {\bibfnamefont {V.~S.}\ \bibnamefont {Bagnato}},
  \bibinfo {author} {\bibfnamefont {S.}~\bibnamefont {Zilio}}, \ and\ \bibinfo
  {author} {\bibfnamefont {P.~S.}\ \bibnamefont {Julienne}},\ }\href {\doibase
  10.1103/RevModPhys.71.1} {\bibfield  {journal} {\bibinfo  {journal} {Rev.
  Mod. Phys.}\ }\textbf {\bibinfo {volume} {71}},\ \bibinfo {pages} {1}
  (\bibinfo {year} {1999})}\BibitemShut {NoStop}%
\bibitem [{\citenamefont {Anderson}\ \emph {et~al.}(1995)\citenamefont
  {Anderson}, \citenamefont {Ensher}, \citenamefont {Matthews}, \citenamefont
  {Wieman},\ and\ \citenamefont {Cornell}}]{AndersonScience1995}%
  \BibitemOpen
  \bibfield  {author} {\bibinfo {author} {\bibfnamefont {M.~H.}\ \bibnamefont
  {Anderson}}, \bibinfo {author} {\bibfnamefont {J.~R.}\ \bibnamefont
  {Ensher}}, \bibinfo {author} {\bibfnamefont {M.~R.}\ \bibnamefont
  {Matthews}}, \bibinfo {author} {\bibfnamefont {C.~E.}\ \bibnamefont
  {Wieman}}, \ and\ \bibinfo {author} {\bibfnamefont {E.~A.}\ \bibnamefont
  {Cornell}},\ }\href {\doibase 10.1126/science.269.5221.198} {\bibfield
  {journal} {\bibinfo  {journal} {Science}\ }\textbf {\bibinfo {volume}
  {269}},\ \bibinfo {pages} {198} (\bibinfo {year} {1995})},\ \Eprint
  {http://arxiv.org/abs/http://www.sciencemag.org/content/269/5221/198.full.pdf}
  {http://www.sciencemag.org/content/269/5221/198.full.pdf} \BibitemShut
  {NoStop}%
\bibitem [{\citenamefont {Davis}\ \emph {et~al.}(1995)\citenamefont {Davis},
  \citenamefont {Mewes}, \citenamefont {Andrews}, \citenamefont {van Druten},
  \citenamefont {Durfee}, \citenamefont {Kurn},\ and\ \citenamefont
  {Ketterle}}]{DavisPRL1995}%
  \BibitemOpen
  \bibfield  {author} {\bibinfo {author} {\bibfnamefont {K.~B.}\ \bibnamefont
  {Davis}}, \bibinfo {author} {\bibfnamefont {M.~O.}\ \bibnamefont {Mewes}},
  \bibinfo {author} {\bibfnamefont {M.~R.}\ \bibnamefont {Andrews}}, \bibinfo
  {author} {\bibfnamefont {N.~J.}\ \bibnamefont {van Druten}}, \bibinfo
  {author} {\bibfnamefont {D.~S.}\ \bibnamefont {Durfee}}, \bibinfo {author}
  {\bibfnamefont {D.~M.}\ \bibnamefont {Kurn}}, \ and\ \bibinfo {author}
  {\bibfnamefont {W.}~\bibnamefont {Ketterle}},\ }\href {\doibase
  10.1103/PhysRevLett.75.3969} {\bibfield  {journal} {\bibinfo  {journal}
  {Phys. Rev. Lett.}\ }\textbf {\bibinfo {volume} {75}},\ \bibinfo {pages}
  {3969} (\bibinfo {year} {1995})}\BibitemShut {NoStop}%
\bibitem [{\citenamefont {Takasu}\ \emph {et~al.}(2003)\citenamefont {Takasu},
  \citenamefont {Maki}, \citenamefont {Komori}, \citenamefont {Takano},
  \citenamefont {Honda}, \citenamefont {Kumakura}, \citenamefont {Yabuzaki},\
  and\ \citenamefont {Takahashi}}]{TakusuPRL2003}%
  \BibitemOpen
  \bibfield  {author} {\bibinfo {author} {\bibfnamefont {Y.}~\bibnamefont
  {Takasu}}, \bibinfo {author} {\bibfnamefont {K.}~\bibnamefont {Maki}},
  \bibinfo {author} {\bibfnamefont {K.}~\bibnamefont {Komori}}, \bibinfo
  {author} {\bibfnamefont {T.}~\bibnamefont {Takano}}, \bibinfo {author}
  {\bibfnamefont {K.}~\bibnamefont {Honda}}, \bibinfo {author} {\bibfnamefont
  {M.}~\bibnamefont {Kumakura}}, \bibinfo {author} {\bibfnamefont
  {T.}~\bibnamefont {Yabuzaki}}, \ and\ \bibinfo {author} {\bibfnamefont
  {Y.}~\bibnamefont {Takahashi}},\ }\href {\doibase
  10.1103/PhysRevLett.91.040404} {\bibfield  {journal} {\bibinfo  {journal}
  {Phys. Rev. Lett.}\ }\textbf {\bibinfo {volume} {91}},\ \bibinfo {pages}
  {040404} (\bibinfo {year} {2003})}\BibitemShut {NoStop}%
\bibitem [{\citenamefont {Kishimoto}\ \emph {et~al.}(2009)\citenamefont
  {Kishimoto}, \citenamefont {Kobayashi}, \citenamefont {Noda}, \citenamefont
  {Aikawa}, \citenamefont {Ueda},\ and\ \citenamefont
  {Inouye}}]{KishimotoPRA2009}%
  \BibitemOpen
  \bibfield  {author} {\bibinfo {author} {\bibfnamefont {T.}~\bibnamefont
  {Kishimoto}}, \bibinfo {author} {\bibfnamefont {J.}~\bibnamefont
  {Kobayashi}}, \bibinfo {author} {\bibfnamefont {K.}~\bibnamefont {Noda}},
  \bibinfo {author} {\bibfnamefont {K.}~\bibnamefont {Aikawa}}, \bibinfo
  {author} {\bibfnamefont {M.}~\bibnamefont {Ueda}}, \ and\ \bibinfo {author}
  {\bibfnamefont {S.}~\bibnamefont {Inouye}},\ }\href {\doibase
  10.1103/PhysRevA.79.031602} {\bibfield  {journal} {\bibinfo  {journal} {Phys.
  Rev. A}\ }\textbf {\bibinfo {volume} {79}},\ \bibinfo {pages} {031602}
  (\bibinfo {year} {2009})}\BibitemShut {NoStop}%
\bibitem [{Note2()}]{Note2}%
  \BibitemOpen
  \bibinfo {note} {For lighter species such as $^7$Li \cite {BradleyPRL1997}
  and metastable He \cite {RobertScience2001,SantosPRL2001}, slightly smaller
  scattering lengths can be used because of the inverse scaling of the
  collision rate with mass.}\BibitemShut {Stop}%
\bibitem [{\citenamefont {Bedaque}\ \emph {et~al.}(2000)\citenamefont
  {Bedaque}, \citenamefont {Braaten},\ and\ \citenamefont
  {Hammer}}]{BedaquePRL2000}%
  \BibitemOpen
  \bibfield  {author} {\bibinfo {author} {\bibfnamefont {P.~F.}\ \bibnamefont
  {Bedaque}}, \bibinfo {author} {\bibfnamefont {E.}~\bibnamefont {Braaten}}, \
  and\ \bibinfo {author} {\bibfnamefont {H.-W.}\ \bibnamefont {Hammer}},\
  }\href {\doibase 10.1103/PhysRevLett.85.908} {\bibfield  {journal} {\bibinfo
  {journal} {Phys. Rev. Lett.}\ }\textbf {\bibinfo {volume} {85}},\ \bibinfo
  {pages} {908} (\bibinfo {year} {2000})}\BibitemShut {NoStop}%
\bibitem [{\citenamefont {Cornish}\ \emph {et~al.}(2000)\citenamefont
  {Cornish}, \citenamefont {Claussen}, \citenamefont {Roberts}, \citenamefont
  {Cornell},\ and\ \citenamefont {Wieman}}]{CornishPRL2000}%
  \BibitemOpen
  \bibfield  {author} {\bibinfo {author} {\bibfnamefont {S.~L.}\ \bibnamefont
  {Cornish}}, \bibinfo {author} {\bibfnamefont {N.~R.}\ \bibnamefont
  {Claussen}}, \bibinfo {author} {\bibfnamefont {J.~L.}\ \bibnamefont
  {Roberts}}, \bibinfo {author} {\bibfnamefont {E.~A.}\ \bibnamefont
  {Cornell}}, \ and\ \bibinfo {author} {\bibfnamefont {C.~E.}\ \bibnamefont
  {Wieman}},\ }\href {\doibase 10.1103/PhysRevLett.85.1795} {\bibfield
  {journal} {\bibinfo  {journal} {Phys. Rev. Lett.}\ }\textbf {\bibinfo
  {volume} {85}},\ \bibinfo {pages} {1795} (\bibinfo {year}
  {2000})}\BibitemShut {NoStop}%
\bibitem [{\citenamefont {Strecker}\ \emph {et~al.}(2002)\citenamefont
  {Strecker}, \citenamefont {Partridge}, \citenamefont {Truscott},\ and\
  \citenamefont {Hulet}}]{StreckerNature2002}%
  \BibitemOpen
  \bibfield  {author} {\bibinfo {author} {\bibfnamefont {K.~E.}\ \bibnamefont
  {Strecker}}, \bibinfo {author} {\bibfnamefont {G.~B.}\ \bibnamefont
  {Partridge}}, \bibinfo {author} {\bibfnamefont {A.~G.}\ \bibnamefont
  {Truscott}}, \ and\ \bibinfo {author} {\bibfnamefont {R.~G.}\ \bibnamefont
  {Hulet}},\ }\href {http://dx.doi.org/10.1038/nature747} {\bibfield  {journal}
  {\bibinfo  {journal} {Nature}\ }\textbf {\bibinfo {volume} {417}},\ \bibinfo
  {pages} {150} (\bibinfo {year} {2002})}\BibitemShut {NoStop}%
\bibitem [{\citenamefont {Weber}\ \emph {et~al.}(2003)\citenamefont {Weber},
  \citenamefont {Herbig}, \citenamefont {Mark}, \citenamefont {N\"agerl},\ and\
  \citenamefont {Grimm}}]{WeberScience2003}%
  \BibitemOpen
  \bibfield  {author} {\bibinfo {author} {\bibfnamefont {T.}~\bibnamefont
  {Weber}}, \bibinfo {author} {\bibfnamefont {J.}~\bibnamefont {Herbig}},
  \bibinfo {author} {\bibfnamefont {M.}~\bibnamefont {Mark}}, \bibinfo {author}
  {\bibfnamefont {H.-C.}\ \bibnamefont {N\"agerl}}, \ and\ \bibinfo {author}
  {\bibfnamefont {R.}~\bibnamefont {Grimm}},\ }\href {\doibase
  10.1126/science.1079699} {\bibfield  {journal} {\bibinfo  {journal}
  {Science}\ }\textbf {\bibinfo {volume} {299}},\ \bibinfo {pages} {232}
  (\bibinfo {year} {2003})},\ \Eprint
  {http://arxiv.org/abs/http://www.sciencemag.org/content/299/5604/232.full.pdf}
  {http://www.sciencemag.org/content/299/5604/232.full.pdf} \BibitemShut
  {NoStop}%
\bibitem [{\citenamefont {Landini}\ \emph {et~al.}(2012)\citenamefont
  {Landini}, \citenamefont {Roy}, \citenamefont {Roati}, \citenamefont
  {Simoni}, \citenamefont {Inguscio}, \citenamefont {Modugno},\ and\
  \citenamefont {Fattori}}]{LandiniPRA2012}%
  \BibitemOpen
  \bibfield  {author} {\bibinfo {author} {\bibfnamefont {M.}~\bibnamefont
  {Landini}}, \bibinfo {author} {\bibfnamefont {S.}~\bibnamefont {Roy}},
  \bibinfo {author} {\bibfnamefont {G.}~\bibnamefont {Roati}}, \bibinfo
  {author} {\bibfnamefont {A.}~\bibnamefont {Simoni}}, \bibinfo {author}
  {\bibfnamefont {M.}~\bibnamefont {Inguscio}}, \bibinfo {author}
  {\bibfnamefont {G.}~\bibnamefont {Modugno}}, \ and\ \bibinfo {author}
  {\bibfnamefont {M.}~\bibnamefont {Fattori}},\ }\href {\doibase
  10.1103/PhysRevA.86.033421} {\bibfield  {journal} {\bibinfo  {journal} {Phys.
  Rev. A}\ }\textbf {\bibinfo {volume} {86}},\ \bibinfo {pages} {033421}
  (\bibinfo {year} {2012})}\BibitemShut {NoStop}%
\bibitem [{\citenamefont {Roberts}\ \emph {et~al.}(2000)\citenamefont
  {Roberts}, \citenamefont {Claussen}, \citenamefont {Cornish},\ and\
  \citenamefont {Wieman}}]{RobertsPRL2000}%
  \BibitemOpen
  \bibfield  {author} {\bibinfo {author} {\bibfnamefont {J.~L.}\ \bibnamefont
  {Roberts}}, \bibinfo {author} {\bibfnamefont {N.~R.}\ \bibnamefont
  {Claussen}}, \bibinfo {author} {\bibfnamefont {S.~L.}\ \bibnamefont
  {Cornish}}, \ and\ \bibinfo {author} {\bibfnamefont {C.~E.}\ \bibnamefont
  {Wieman}},\ }\href {\doibase 10.1103/PhysRevLett.85.728} {\bibfield
  {journal} {\bibinfo  {journal} {Phys. Rev. Lett.}\ }\textbf {\bibinfo
  {volume} {85}},\ \bibinfo {pages} {728} (\bibinfo {year} {2000})}\BibitemShut
  {NoStop}%
\bibitem [{\citenamefont {Marchant}\ \emph {et~al.}(2012)\citenamefont
  {Marchant}, \citenamefont {H\"andel}, \citenamefont {Hopkins}, \citenamefont
  {Wiles},\ and\ \citenamefont {Cornish}}]{MarchantPRA2012}%
  \BibitemOpen
  \bibfield  {author} {\bibinfo {author} {\bibfnamefont {A.~L.}\ \bibnamefont
  {Marchant}}, \bibinfo {author} {\bibfnamefont {S.}~\bibnamefont {H\"andel}},
  \bibinfo {author} {\bibfnamefont {S.~A.}\ \bibnamefont {Hopkins}}, \bibinfo
  {author} {\bibfnamefont {T.~P.}\ \bibnamefont {Wiles}}, \ and\ \bibinfo
  {author} {\bibfnamefont {S.~L.}\ \bibnamefont {Cornish}},\ }\href {\doibase
  10.1103/PhysRevA.85.053647} {\bibfield  {journal} {\bibinfo  {journal} {Phys.
  Rev. A}\ }\textbf {\bibinfo {volume} {85}},\ \bibinfo {pages} {053647}
  (\bibinfo {year} {2012})}\BibitemShut {NoStop}%
\bibitem [{\citenamefont {Efimov}(1970)}]{EfimovPLB1970}%
  \BibitemOpen
  \bibfield  {author} {\bibinfo {author} {\bibfnamefont {V.}~\bibnamefont
  {Efimov}},\ }\href {\doibase http://dx.doi.org/10.1016/0370-2693(70)90349-7}
  {\bibfield  {journal} {\bibinfo  {journal} {Physics Letters B}\ }\textbf
  {\bibinfo {volume} {33}},\ \bibinfo {pages} {563 } (\bibinfo {year}
  {1970})}\BibitemShut {NoStop}%
\bibitem [{\citenamefont {Braaten}\ and\ \citenamefont
  {Hammer}(2006)}]{BraatenPhysRep2006}%
  \BibitemOpen
  \bibfield  {author} {\bibinfo {author} {\bibfnamefont {E.}~\bibnamefont
  {Braaten}}\ and\ \bibinfo {author} {\bibfnamefont {H.-W.}\ \bibnamefont
  {Hammer}},\ }\href {\doibase http://dx.doi.org/10.1016/j.physrep.2006.03.001}
  {\bibfield  {journal} {\bibinfo  {journal} {Physics Reports}\ }\textbf
  {\bibinfo {volume} {428}},\ \bibinfo {pages} {259 } (\bibinfo {year}
  {2006})}\BibitemShut {NoStop}%
\bibitem [{\citenamefont {Kraemer}\ \emph {et~al.}(2006)\citenamefont
  {Kraemer}, \citenamefont {Mark}, \citenamefont {Waldburger}, \citenamefont
  {Danzl}, \citenamefont {Chin}, \citenamefont {Engeser}, \citenamefont
  {Lange}, \citenamefont {Pilch}, \citenamefont {Jaakkola}, \citenamefont
  {N\"{a}gerl},\ and\ \citenamefont {Grimm}}]{Kraemer:2006}%
  \BibitemOpen
  \bibfield  {author} {\bibinfo {author} {\bibfnamefont {T.}~\bibnamefont
  {Kraemer}}, \bibinfo {author} {\bibfnamefont {M.}~\bibnamefont {Mark}},
  \bibinfo {author} {\bibfnamefont {P.}~\bibnamefont {Waldburger}}, \bibinfo
  {author} {\bibfnamefont {J.~G.}\ \bibnamefont {Danzl}}, \bibinfo {author}
  {\bibfnamefont {C.}~\bibnamefont {Chin}}, \bibinfo {author} {\bibfnamefont
  {B.}~\bibnamefont {Engeser}}, \bibinfo {author} {\bibfnamefont {A.~D.}\
  \bibnamefont {Lange}}, \bibinfo {author} {\bibfnamefont {K.}~\bibnamefont
  {Pilch}}, \bibinfo {author} {\bibfnamefont {A.}~\bibnamefont {Jaakkola}},
  \bibinfo {author} {\bibfnamefont {H.~C.}\ \bibnamefont {N\"{a}gerl}}, \ and\
  \bibinfo {author} {\bibfnamefont {R.}~\bibnamefont {Grimm}},\ }\href@noop {}
  {\bibfield  {journal} {\bibinfo  {journal} {Nature}\ }\textbf {\bibinfo
  {volume} {440}},\ \bibinfo {pages} {315} (\bibinfo {year}
  {2006})}\BibitemShut {NoStop}%
\bibitem [{\citenamefont {Berninger}\ \emph {et~al.}(2011)\citenamefont
  {Berninger}, \citenamefont {Zenesini}, \citenamefont {Huang}, \citenamefont
  {Harm}, \citenamefont {N\"agerl}, \citenamefont {Ferlaino}, \citenamefont
  {Grimm}, \citenamefont {Julienne},\ and\ \citenamefont
  {Hutson}}]{Berninger:Efimov:2011}%
  \BibitemOpen
  \bibfield  {author} {\bibinfo {author} {\bibfnamefont {M.}~\bibnamefont
  {Berninger}}, \bibinfo {author} {\bibfnamefont {A.}~\bibnamefont {Zenesini}},
  \bibinfo {author} {\bibfnamefont {B.}~\bibnamefont {Huang}}, \bibinfo
  {author} {\bibfnamefont {W.}~\bibnamefont {Harm}}, \bibinfo {author}
  {\bibfnamefont {H.-C.}\ \bibnamefont {N\"agerl}}, \bibinfo {author}
  {\bibfnamefont {F.}~\bibnamefont {Ferlaino}}, \bibinfo {author}
  {\bibfnamefont {R.}~\bibnamefont {Grimm}}, \bibinfo {author} {\bibfnamefont
  {P.~S.}\ \bibnamefont {Julienne}}, \ and\ \bibinfo {author} {\bibfnamefont
  {J.~M.}\ \bibnamefont {Hutson}},\ }\href@noop {} {\bibfield  {journal}
  {\bibinfo  {journal} {Phys. Rev. Lett.}\ }\textbf {\bibinfo {volume} {107}},\
  \bibinfo {pages} {120401} (\bibinfo {year} {2011})}\BibitemShut {NoStop}%
\bibitem [{\citenamefont {Myatt}\ \emph {et~al.}(1997)\citenamefont {Myatt},
  \citenamefont {Burt}, \citenamefont {Ghrist}, \citenamefont {Cornell},\ and\
  \citenamefont {Wieman}}]{MyattPRL1997}%
  \BibitemOpen
  \bibfield  {author} {\bibinfo {author} {\bibfnamefont {C.~J.}\ \bibnamefont
  {Myatt}}, \bibinfo {author} {\bibfnamefont {E.~A.}\ \bibnamefont {Burt}},
  \bibinfo {author} {\bibfnamefont {R.~W.}\ \bibnamefont {Ghrist}}, \bibinfo
  {author} {\bibfnamefont {E.~A.}\ \bibnamefont {Cornell}}, \ and\ \bibinfo
  {author} {\bibfnamefont {C.~E.}\ \bibnamefont {Wieman}},\ }\href {\doibase
  10.1103/PhysRevLett.78.586} {\bibfield  {journal} {\bibinfo  {journal} {Phys.
  Rev. Lett.}\ }\textbf {\bibinfo {volume} {78}},\ \bibinfo {pages} {586}
  (\bibinfo {year} {1997})}\BibitemShut {NoStop}%
\bibitem [{\citenamefont {Modugno}\ \emph {et~al.}(2002)\citenamefont
  {Modugno}, \citenamefont {Modugno}, \citenamefont {Riboli}, \citenamefont
  {Roati},\ and\ \citenamefont {Inguscio}}]{ModugnoPRL2002}%
  \BibitemOpen
  \bibfield  {author} {\bibinfo {author} {\bibfnamefont {G.}~\bibnamefont
  {Modugno}}, \bibinfo {author} {\bibfnamefont {M.}~\bibnamefont {Modugno}},
  \bibinfo {author} {\bibfnamefont {F.}~\bibnamefont {Riboli}}, \bibinfo
  {author} {\bibfnamefont {G.}~\bibnamefont {Roati}}, \ and\ \bibinfo {author}
  {\bibfnamefont {M.}~\bibnamefont {Inguscio}},\ }\href {\doibase
  10.1103/PhysRevLett.89.190404} {\bibfield  {journal} {\bibinfo  {journal}
  {Phys. Rev. Lett.}\ }\textbf {\bibinfo {volume} {89}},\ \bibinfo {pages}
  {190404} (\bibinfo {year} {2002})}\BibitemShut {NoStop}%
\bibitem [{\citenamefont {Hadzibabic}\ \emph {et~al.}(2002)\citenamefont
  {Hadzibabic}, \citenamefont {Stan}, \citenamefont {Dieckmann}, \citenamefont
  {Gupta}, \citenamefont {Zwierlein}, \citenamefont {G\"orlitz},\ and\
  \citenamefont {Ketterle}}]{HadzibabicPRL2002}%
  \BibitemOpen
  \bibfield  {author} {\bibinfo {author} {\bibfnamefont {Z.}~\bibnamefont
  {Hadzibabic}}, \bibinfo {author} {\bibfnamefont {C.~A.}\ \bibnamefont
  {Stan}}, \bibinfo {author} {\bibfnamefont {K.}~\bibnamefont {Dieckmann}},
  \bibinfo {author} {\bibfnamefont {S.}~\bibnamefont {Gupta}}, \bibinfo
  {author} {\bibfnamefont {M.~W.}\ \bibnamefont {Zwierlein}}, \bibinfo {author}
  {\bibfnamefont {A.}~\bibnamefont {G\"orlitz}}, \ and\ \bibinfo {author}
  {\bibfnamefont {W.}~\bibnamefont {Ketterle}},\ }\href {\doibase
  10.1103/PhysRevLett.88.160401} {\bibfield  {journal} {\bibinfo  {journal}
  {Phys. Rev. Lett.}\ }\textbf {\bibinfo {volume} {88}},\ \bibinfo {pages}
  {160401} (\bibinfo {year} {2002})}\BibitemShut {NoStop}%
\bibitem [{\citenamefont {Roati}\ \emph {et~al.}(2002)\citenamefont {Roati},
  \citenamefont {Riboli}, \citenamefont {Modugno},\ and\ \citenamefont
  {Inguscio}}]{RoatiPRL2002}%
  \BibitemOpen
  \bibfield  {author} {\bibinfo {author} {\bibfnamefont {G.}~\bibnamefont
  {Roati}}, \bibinfo {author} {\bibfnamefont {F.}~\bibnamefont {Riboli}},
  \bibinfo {author} {\bibfnamefont {G.}~\bibnamefont {Modugno}}, \ and\
  \bibinfo {author} {\bibfnamefont {M.}~\bibnamefont {Inguscio}},\ }\href
  {\doibase 10.1103/PhysRevLett.89.150403} {\bibfield  {journal} {\bibinfo
  {journal} {Phys. Rev. Lett.}\ }\textbf {\bibinfo {volume} {89}},\ \bibinfo
  {pages} {150403} (\bibinfo {year} {2002})}\BibitemShut {NoStop}%
\bibitem [{\citenamefont {Silber}\ \emph {et~al.}(2005)\citenamefont {Silber},
  \citenamefont {G\"unther}, \citenamefont {Marzok}, \citenamefont {Deh},
  \citenamefont {Courteille},\ and\ \citenamefont
  {Zimmermann}}]{SilberPRL2005}%
  \BibitemOpen
  \bibfield  {author} {\bibinfo {author} {\bibfnamefont {C.}~\bibnamefont
  {Silber}}, \bibinfo {author} {\bibfnamefont {S.}~\bibnamefont {G\"unther}},
  \bibinfo {author} {\bibfnamefont {C.}~\bibnamefont {Marzok}}, \bibinfo
  {author} {\bibfnamefont {B.}~\bibnamefont {Deh}}, \bibinfo {author}
  {\bibfnamefont {P.~W.}\ \bibnamefont {Courteille}}, \ and\ \bibinfo {author}
  {\bibfnamefont {C.}~\bibnamefont {Zimmermann}},\ }\href {\doibase
  10.1103/PhysRevLett.95.170408} {\bibfield  {journal} {\bibinfo  {journal}
  {Phys. Rev. Lett.}\ }\textbf {\bibinfo {volume} {95}},\ \bibinfo {pages}
  {170408} (\bibinfo {year} {2005})}\BibitemShut {NoStop}%
\bibitem [{\citenamefont {Papp}\ \emph {et~al.}(2008)\citenamefont {Papp},
  \citenamefont {Pino},\ and\ \citenamefont {Wieman}}]{PappPRL2008}%
  \BibitemOpen
  \bibfield  {author} {\bibinfo {author} {\bibfnamefont {S.~B.}\ \bibnamefont
  {Papp}}, \bibinfo {author} {\bibfnamefont {J.~M.}\ \bibnamefont {Pino}}, \
  and\ \bibinfo {author} {\bibfnamefont {C.~E.}\ \bibnamefont {Wieman}},\
  }\href {\doibase 10.1103/PhysRevLett.101.040402} {\bibfield  {journal}
  {\bibinfo  {journal} {Phys. Rev. Lett.}\ }\textbf {\bibinfo {volume} {101}},\
  \bibinfo {pages} {040402} (\bibinfo {year} {2008})}\BibitemShut {NoStop}%
\bibitem [{\citenamefont {Cho}\ \emph {et~al.}(2013)\citenamefont {Cho},
  \citenamefont {McCarron}, \citenamefont {K\"oppinger}, \citenamefont
  {Jenkin}, \citenamefont {Butler}, \citenamefont {Julienne}, \citenamefont
  {Blackley}, \citenamefont {Le~Sueur}, \citenamefont {Hutson},\ and\
  \citenamefont {Cornish}}]{ChoPRA2013}%
  \BibitemOpen
  \bibfield  {author} {\bibinfo {author} {\bibfnamefont {H.-W.}\ \bibnamefont
  {Cho}}, \bibinfo {author} {\bibfnamefont {D.~J.}\ \bibnamefont {McCarron}},
  \bibinfo {author} {\bibfnamefont {M.~P.}\ \bibnamefont {K\"oppinger}},
  \bibinfo {author} {\bibfnamefont {D.~L.}\ \bibnamefont {Jenkin}}, \bibinfo
  {author} {\bibfnamefont {K.~L.}\ \bibnamefont {Butler}}, \bibinfo {author}
  {\bibfnamefont {P.~S.}\ \bibnamefont {Julienne}}, \bibinfo {author}
  {\bibfnamefont {C.~L.}\ \bibnamefont {Blackley}}, \bibinfo {author}
  {\bibfnamefont {C.~R.}\ \bibnamefont {Le~Sueur}}, \bibinfo {author}
  {\bibfnamefont {J.~M.}\ \bibnamefont {Hutson}}, \ and\ \bibinfo {author}
  {\bibfnamefont {S.~L.}\ \bibnamefont {Cornish}},\ }\href {\doibase
  10.1103/PhysRevA.87.010703} {\bibfield  {journal} {\bibinfo  {journal} {Phys.
  Rev. A}\ }\textbf {\bibinfo {volume} {87}},\ \bibinfo {pages} {010703}
  (\bibinfo {year} {2013})}\BibitemShut {NoStop}%
\bibitem [{\citenamefont {Riboli}\ and\ \citenamefont
  {Modugno}(2002)}]{RiboliPRA2002}%
  \BibitemOpen
  \bibfield  {author} {\bibinfo {author} {\bibfnamefont {F.}~\bibnamefont
  {Riboli}}\ and\ \bibinfo {author} {\bibfnamefont {M.}~\bibnamefont
  {Modugno}},\ }\href {\doibase 10.1103/PhysRevA.65.063614} {\bibfield
  {journal} {\bibinfo  {journal} {Phys. Rev. A}\ }\textbf {\bibinfo {volume}
  {65}},\ \bibinfo {pages} {063614} (\bibinfo {year} {2002})}\BibitemShut
  {NoStop}%
\bibitem [{\citenamefont {Damski}\ \emph {et~al.}(2003)\citenamefont {Damski},
  \citenamefont {Santos}, \citenamefont {Tiemann}, \citenamefont {Lewenstein},
  \citenamefont {Kotochigova}, \citenamefont {Julienne},\ and\ \citenamefont
  {Zoller}}]{DamskiPRL2003}%
  \BibitemOpen
  \bibfield  {author} {\bibinfo {author} {\bibfnamefont {B.}~\bibnamefont
  {Damski}}, \bibinfo {author} {\bibfnamefont {L.}~\bibnamefont {Santos}},
  \bibinfo {author} {\bibfnamefont {E.}~\bibnamefont {Tiemann}}, \bibinfo
  {author} {\bibfnamefont {M.}~\bibnamefont {Lewenstein}}, \bibinfo {author}
  {\bibfnamefont {S.}~\bibnamefont {Kotochigova}}, \bibinfo {author}
  {\bibfnamefont {P.}~\bibnamefont {Julienne}}, \ and\ \bibinfo {author}
  {\bibfnamefont {P.}~\bibnamefont {Zoller}},\ }\href {\doibase
  10.1103/PhysRevLett.90.110401} {\bibfield  {journal} {\bibinfo  {journal}
  {Phys. Rev. Lett.}\ }\textbf {\bibinfo {volume} {90}},\ \bibinfo {pages}
  {110401} (\bibinfo {year} {2003})}\BibitemShut {NoStop}%
\bibitem [{\citenamefont {Hung}\ \emph {et~al.}(2008)\citenamefont {Hung},
  \citenamefont {Zhang}, \citenamefont {Gemelke},\ and\ \citenamefont
  {Chin}}]{HungPRA2008}%
  \BibitemOpen
  \bibfield  {author} {\bibinfo {author} {\bibfnamefont {C.-L.}\ \bibnamefont
  {Hung}}, \bibinfo {author} {\bibfnamefont {X.}~\bibnamefont {Zhang}},
  \bibinfo {author} {\bibfnamefont {N.}~\bibnamefont {Gemelke}}, \ and\
  \bibinfo {author} {\bibfnamefont {C.}~\bibnamefont {Chin}},\ }\href {\doibase
  10.1103/PhysRevA.78.011604} {\bibfield  {journal} {\bibinfo  {journal} {Phys.
  Rev. A}\ }\textbf {\bibinfo {volume} {78}},\ \bibinfo {pages} {011604}
  (\bibinfo {year} {2008})}\BibitemShut {NoStop}%
\bibitem [{\citenamefont {McCarron}\ \emph {et~al.}(2011)\citenamefont
  {McCarron}, \citenamefont {Cho}, \citenamefont {Jenkin}, \citenamefont
  {K\"oppinger},\ and\ \citenamefont {Cornish}}]{McCarronPRA2011}%
  \BibitemOpen
  \bibfield  {author} {\bibinfo {author} {\bibfnamefont {D.~J.}\ \bibnamefont
  {McCarron}}, \bibinfo {author} {\bibfnamefont {H.~W.}\ \bibnamefont {Cho}},
  \bibinfo {author} {\bibfnamefont {D.~L.}\ \bibnamefont {Jenkin}}, \bibinfo
  {author} {\bibfnamefont {M.~P.}\ \bibnamefont {K\"oppinger}}, \ and\ \bibinfo
  {author} {\bibfnamefont {S.~L.}\ \bibnamefont {Cornish}},\ }\href {\doibase
  10.1103/PhysRevA.84.011603} {\bibfield  {journal} {\bibinfo  {journal} {Phys.
  Rev. A}\ }\textbf {\bibinfo {volume} {84}},\ \bibinfo {pages} {011603}
  (\bibinfo {year} {2011})}\BibitemShut {NoStop}%
\bibitem [{\citenamefont {Roati}\ \emph {et~al.}(2007)\citenamefont {Roati},
  \citenamefont {Zaccanti}, \citenamefont {D'Errico}, \citenamefont {Catani},
  \citenamefont {Modugno}, \citenamefont {Simoni}, \citenamefont {Inguscio},\
  and\ \citenamefont {Modugno}}]{RoatiPRL2007}%
  \BibitemOpen
  \bibfield  {author} {\bibinfo {author} {\bibfnamefont {G.}~\bibnamefont
  {Roati}}, \bibinfo {author} {\bibfnamefont {M.}~\bibnamefont {Zaccanti}},
  \bibinfo {author} {\bibfnamefont {C.}~\bibnamefont {D'Errico}}, \bibinfo
  {author} {\bibfnamefont {J.}~\bibnamefont {Catani}}, \bibinfo {author}
  {\bibfnamefont {M.}~\bibnamefont {Modugno}}, \bibinfo {author} {\bibfnamefont
  {A.}~\bibnamefont {Simoni}}, \bibinfo {author} {\bibfnamefont
  {M.}~\bibnamefont {Inguscio}}, \ and\ \bibinfo {author} {\bibfnamefont
  {G.}~\bibnamefont {Modugno}},\ }\href {\doibase
  10.1103/PhysRevLett.99.010403} {\bibfield  {journal} {\bibinfo  {journal}
  {Phys. Rev. Lett.}\ }\textbf {\bibinfo {volume} {99}},\ \bibinfo {pages}
  {010403} (\bibinfo {year} {2007})}\BibitemShut {NoStop}%
\bibitem [{\citenamefont {Burt}\ \emph {et~al.}(1997)\citenamefont {Burt},
  \citenamefont {Ghrist}, \citenamefont {Myatt}, \citenamefont {Holland},
  \citenamefont {Cornell},\ and\ \citenamefont {Wieman}}]{Burt:1997}%
  \BibitemOpen
  \bibfield  {author} {\bibinfo {author} {\bibfnamefont {E.~A.}\ \bibnamefont
  {Burt}}, \bibinfo {author} {\bibfnamefont {R.~W.}\ \bibnamefont {Ghrist}},
  \bibinfo {author} {\bibfnamefont {C.~J.}\ \bibnamefont {Myatt}}, \bibinfo
  {author} {\bibfnamefont {M.~J.}\ \bibnamefont {Holland}}, \bibinfo {author}
  {\bibfnamefont {E.~A.}\ \bibnamefont {Cornell}}, \ and\ \bibinfo {author}
  {\bibfnamefont {C.~E.}\ \bibnamefont {Wieman}},\ }\href {\doibase
  10.1103/PhysRevLett.79.337} {\bibfield  {journal} {\bibinfo  {journal} {Phys.
  Rev. Lett.}\ }\textbf {\bibinfo {volume} {79}},\ \bibinfo {pages} {337}
  (\bibinfo {year} {1997})}\BibitemShut {NoStop}%
\bibitem [{\citenamefont {Chin}\ \emph {et~al.}(2005)\citenamefont {Chin},
  \citenamefont {Kraemer}, \citenamefont {Mark}, \citenamefont {Herbig},
  \citenamefont {Waldburger}, \citenamefont {N\"agerl},\ and\ \citenamefont
  {Grimm}}]{ChinPRL2005}%
  \BibitemOpen
  \bibfield  {author} {\bibinfo {author} {\bibfnamefont {C.}~\bibnamefont
  {Chin}}, \bibinfo {author} {\bibfnamefont {T.}~\bibnamefont {Kraemer}},
  \bibinfo {author} {\bibfnamefont {M.}~\bibnamefont {Mark}}, \bibinfo {author}
  {\bibfnamefont {J.}~\bibnamefont {Herbig}}, \bibinfo {author} {\bibfnamefont
  {P.}~\bibnamefont {Waldburger}}, \bibinfo {author} {\bibfnamefont {H.-C.}\
  \bibnamefont {N\"agerl}}, \ and\ \bibinfo {author} {\bibfnamefont
  {R.}~\bibnamefont {Grimm}},\ }\href {\doibase 10.1103/PhysRevLett.94.123201}
  {\bibfield  {journal} {\bibinfo  {journal} {Phys. Rev. Lett.}\ }\textbf
  {\bibinfo {volume} {94}},\ \bibinfo {pages} {123201} (\bibinfo {year}
  {2005})}\BibitemShut {NoStop}%
\bibitem [{\citenamefont {Herbig}\ \emph {et~al.}(2003)\citenamefont {Herbig},
  \citenamefont {Kraemer}, \citenamefont {Mark}, \citenamefont {Weber},
  \citenamefont {Chin}, \citenamefont {N\"agerl},\ and\ \citenamefont
  {Grimm}}]{HerbigScience2003}%
  \BibitemOpen
  \bibfield  {author} {\bibinfo {author} {\bibfnamefont {J.}~\bibnamefont
  {Herbig}}, \bibinfo {author} {\bibfnamefont {T.}~\bibnamefont {Kraemer}},
  \bibinfo {author} {\bibfnamefont {M.}~\bibnamefont {Mark}}, \bibinfo {author}
  {\bibfnamefont {T.}~\bibnamefont {Weber}}, \bibinfo {author} {\bibfnamefont
  {C.}~\bibnamefont {Chin}}, \bibinfo {author} {\bibfnamefont {H.-C.}\
  \bibnamefont {N\"agerl}}, \ and\ \bibinfo {author} {\bibfnamefont
  {R.}~\bibnamefont {Grimm}},\ }\href {\doibase 10.1126/science.1088876}
  {\bibfield  {journal} {\bibinfo  {journal} {Science}\ }\textbf {\bibinfo
  {volume} {301}},\ \bibinfo {pages} {1510} (\bibinfo {year}
  {2003})}\BibitemShut {NoStop}%
\bibitem [{\citenamefont {Mark}\ \emph {et~al.}(2005)\citenamefont {Mark},
  \citenamefont {Kraemer}, \citenamefont {Herbig}, \citenamefont {Chin},
  \citenamefont {N\"agerl},\ and\ \citenamefont {Grimm}}]{MarkEPL2005}%
  \BibitemOpen
  \bibfield  {author} {\bibinfo {author} {\bibfnamefont {M.}~\bibnamefont
  {Mark}}, \bibinfo {author} {\bibfnamefont {T.}~\bibnamefont {Kraemer}},
  \bibinfo {author} {\bibfnamefont {J.}~\bibnamefont {Herbig}}, \bibinfo
  {author} {\bibfnamefont {C.}~\bibnamefont {Chin}}, \bibinfo {author}
  {\bibfnamefont {H.-C.}\ \bibnamefont {N\"agerl}}, \ and\ \bibinfo {author}
  {\bibfnamefont {R.}~\bibnamefont {Grimm}},\ }\href
  {http://stacks.iop.org/0295-5075/69/i=5/a=706} {\bibfield  {journal}
  {\bibinfo  {journal} {EPL (Europhysics Letters)}\ }\textbf {\bibinfo {volume}
  {69}},\ \bibinfo {pages} {706} (\bibinfo {year} {2005})}\BibitemShut
  {NoStop}%
\bibitem [{\citenamefont {DeMarco}\ and\ \citenamefont
  {Jin}(1999)}]{DeMarcoScience1999}%
  \BibitemOpen
  \bibfield  {author} {\bibinfo {author} {\bibfnamefont {B.}~\bibnamefont
  {DeMarco}}\ and\ \bibinfo {author} {\bibfnamefont {D.~S.}\ \bibnamefont
  {Jin}},\ }\href {\doibase 10.1126/science.285.5434.1703} {\bibfield
  {journal} {\bibinfo  {journal} {Science}\ }\textbf {\bibinfo {volume}
  {285}},\ \bibinfo {pages} {1703} (\bibinfo {year} {1999})},\ \Eprint
  {http://arxiv.org/abs/http://www.sciencemag.org/content/285/5434/1703.full.pdf}
  {http://www.sciencemag.org/content/285/5434/1703.full.pdf} \BibitemShut
  {NoStop}%
\bibitem [{\citenamefont {Bradley}\ \emph {et~al.}(1997)\citenamefont
  {Bradley}, \citenamefont {Sackett},\ and\ \citenamefont
  {Hulet}}]{BradleyPRL1997}%
  \BibitemOpen
  \bibfield  {author} {\bibinfo {author} {\bibfnamefont {C.~C.}\ \bibnamefont
  {Bradley}}, \bibinfo {author} {\bibfnamefont {C.~A.}\ \bibnamefont
  {Sackett}}, \ and\ \bibinfo {author} {\bibfnamefont {R.~G.}\ \bibnamefont
  {Hulet}},\ }\href {\doibase 10.1103/PhysRevLett.78.985} {\bibfield  {journal}
  {\bibinfo  {journal} {Phys. Rev. Lett.}\ }\textbf {\bibinfo {volume} {78}},\
  \bibinfo {pages} {985} (\bibinfo {year} {1997})}\BibitemShut {NoStop}%
\bibitem [{\citenamefont {Robert}\ \emph {et~al.}(2001)\citenamefont {Robert},
  \citenamefont {Sirjean}, \citenamefont {Browaeys}, \citenamefont {Poupard},
  \citenamefont {Nowak}, \citenamefont {Boiron}, \citenamefont {Westbrook},\
  and\ \citenamefont {Aspect}}]{RobertScience2001}%
  \BibitemOpen
  \bibfield  {author} {\bibinfo {author} {\bibfnamefont {A.}~\bibnamefont
  {Robert}}, \bibinfo {author} {\bibfnamefont {O.}~\bibnamefont {Sirjean}},
  \bibinfo {author} {\bibfnamefont {A.}~\bibnamefont {Browaeys}}, \bibinfo
  {author} {\bibfnamefont {J.}~\bibnamefont {Poupard}}, \bibinfo {author}
  {\bibfnamefont {S.}~\bibnamefont {Nowak}}, \bibinfo {author} {\bibfnamefont
  {D.}~\bibnamefont {Boiron}}, \bibinfo {author} {\bibfnamefont {C.~I.}\
  \bibnamefont {Westbrook}}, \ and\ \bibinfo {author} {\bibfnamefont
  {A.}~\bibnamefont {Aspect}},\ }\href {\doibase 10.1126/science.1060622}
  {\bibfield  {journal} {\bibinfo  {journal} {Science}\ }\textbf {\bibinfo
  {volume} {292}},\ \bibinfo {pages} {461} (\bibinfo {year} {2001})},\ \Eprint
  {http://arxiv.org/abs/http://www.sciencemag.org/content/292/5516/461.full.pdf}
  {http://www.sciencemag.org/content/292/5516/461.full.pdf} \BibitemShut
  {NoStop}%
\bibitem [{\citenamefont {Pereira Dos~Santos}\ \emph
  {et~al.}(2001)\citenamefont {Pereira Dos~Santos}, \citenamefont {L\'eonard},
  \citenamefont {Wang}, \citenamefont {Barrelet}, \citenamefont {Perales},
  \citenamefont {Rasel}, \citenamefont {Unnikrishnan}, \citenamefont {Leduc},\
  and\ \citenamefont {Cohen-Tannoudji}}]{SantosPRL2001}%
  \BibitemOpen
  \bibfield  {author} {\bibinfo {author} {\bibfnamefont {F.}~\bibnamefont
  {Pereira Dos~Santos}}, \bibinfo {author} {\bibfnamefont {J.}~\bibnamefont
  {L\'eonard}}, \bibinfo {author} {\bibfnamefont {J.}~\bibnamefont {Wang}},
  \bibinfo {author} {\bibfnamefont {C.~J.}\ \bibnamefont {Barrelet}}, \bibinfo
  {author} {\bibfnamefont {F.}~\bibnamefont {Perales}}, \bibinfo {author}
  {\bibfnamefont {E.}~\bibnamefont {Rasel}}, \bibinfo {author} {\bibfnamefont
  {C.~S.}\ \bibnamefont {Unnikrishnan}}, \bibinfo {author} {\bibfnamefont
  {M.}~\bibnamefont {Leduc}}, \ and\ \bibinfo {author} {\bibfnamefont
  {C.}~\bibnamefont {Cohen-Tannoudji}},\ }\href {\doibase
  10.1103/PhysRevLett.86.3459} {\bibfield  {journal} {\bibinfo  {journal}
  {Phys. Rev. Lett.}\ }\textbf {\bibinfo {volume} {86}},\ \bibinfo {pages}
  {3459} (\bibinfo {year} {2001})}\BibitemShut {NoStop}%
\end{thebibliography}%

\end{document}